\documentclass[12pt]{iopart}
\usepackage{graphicx}
\usepackage{lscape}

\begin{document}
\begin{center}
{\Large\bf Analytical description of the structure of chaos}\\
\vskip 1cm
M. Harsoula, G. Contopoulos and C. Efthymiopoulos\\
Research Center for Astronomy and Applied Mathematics,
Academy of Athens\\
\vskip 1cm
\end{center}

\noindent {\small {\bf Abstract:} We consider analytical formulae
that describe the chaotic regions around the main periodic orbit
$(x=y=0)$ of the H\'{e}non map. Following our previous paper
(Efthymiopoulos, Contopoulos, Katsanikas $2014$) we introduce new
variables $(\xi, \eta)$ in which the product $\xi\eta=c$ (constant)
gives hyperbolic invariant curves. These hyperbolae are mapped by a
canonical transformation $\Phi$ to the plane $(x,y)$, giving "Moser
invariant curves". We find that the series $\Phi$ are convergent up
to a maximum value of $c=c_{max}$. We give estimates of the errors
due to the finite truncation of the series and discuss how these
errors affect the applicability of analytical computations. For
values of the basic parameter $\kappa$ of the H\'{e}non map smaller
than a critical value, there is an island of stability, around a
stable periodic orbit $S$, containing KAM invariant curves. The
Moser curves for $c \leq 0.32$ are completely outside the last KAM
curve around $S$, the curves with $0.32<c<0.41$ intersect the last
KAM curve and the curves with $0.41\leq c< c_{max} \simeq 0.49$ are
completely inside the last KAM curve. All orbits in the chaotic
region around the periodic orbit $(x=y=0)$, although they seem
random, belong to Moser invariant curves, which, therefore define a
"structure of chaos". Orbits starting close and outside the last KAM
curve remain close to it for a stickiness time that is estimated
analytically using the series $\Phi$. We finally calculate  the
periodic orbits that accumulate close to the homoclinic points, i.e.
the points of intersection of the asymptotic curves from $x=y=0$,
exploiting a method based on the self-intersections of the invariant
Moser curves. We find that all the computed periodic orbits are
generated from the stable orbit $S$ for smaller values of the
H\'{e}non parameter $\kappa$, i.e. they are all regular periodic
orbits. }

\section{Introduction}
In a previous paper (Efthymiopoulos et al. 2014) we explored the
applicability of analytical formulae describing the asymptotic
manifolds emanating from unstable periodic orbits in conservative 2D
maps and in Hamiltonian systems of two degrees of freedom. Such
formulae were found by the method of {\it hyperbolic normal form}.
As it was shown by Cherry (1926), Moser (1956, 1958) and Giorgilli
(2001) the method defines a {\it convergent} series transformation
such that the dynamics close to a hyperbolic fixed point takes an
integrable form. The associated integral has the form $\xi\eta=c$,
where $(\xi,\eta)$ are new canonical variables defined from the
original variables by a convergent transformation. The value $c=0$
corresponds to the asymptotic manifolds emanating from the periodic
orbit. The proof of the convergence is based on the fact that the
hyperbolic normal form series do not contain terms with small
divisors, in contrast with the corresponding series around a stable
periodic orbit, which are in general only formal.

Applications of such series in simple mappings were made by da Silva
Ritter et al. (1987), who gave the stable and unstable asymptotic
curves up to a large extent by using high order truncations of the
series. In particular they found a number of homoclinic points
(i.e.intersections of the stable and unstable manifolds of the same
periodic orbit). In the case of continuous Hamiltonian flows, Vieira
and Ozorio de Almeida (1996), and Ozorio de Almeida and Vieira
(1997) attempted to compute homoclinic orbits by purely analytical
means. However, as we have shown in Efthymiopoulos et al. (2014),
the domain of convergence of the hyperbolic normal form in the
Hamiltonian case contains no homoclinic points (see also Bongini et
al. (2001)). The limits of the domain of convergence can be
associated with some singularities of the equations of the
Hamiltonian flow in the complex time domain. However it is possible
to overcome this problem (Efthymiopoulos et al. 2014) by a method of
analytic continuation allowing to extent the computation of the
asymptotic curves to an arbitrarily large length using only series.
This allowed, in turn, the calculation of many homoclinic points. In
general, the hyperbolic normal form lends itself quite conveniently
to the computation of both homoclinic and heteroclinic points (i.e.
intersections of the manifolds of different periodic orbits, see
Contopoulos et al. 2013).

The asymptotic curves make an infinite number of oscillations
passing close to the unstable periodic points but also extending to
large distances from them. Thus, the asymptotic orbits, which start
on these curves, go to large distances but return an infinite number
of times close to the original periodic orbit. The successive points
of intersection of an asymptotic orbit by a surface of section are
distributed in an apparently random way on this surface, belonging
to what is commomly referred to as the chaotic layer (or chaotic
sea) around the unstable periodic orbit. However, the fact that the
successive consequents of an asymptotic orbit can, in principle, be
{\it all} computed by convergent series implies that these points
are not truly randomly distributed, i.e., that there is some
underlying `structure of chaos'. In the same way, the orbits with
initial conditions close to the asymptotic ones also belong to
invariant curves defined by the integral $c$ for values $c\neq 0$.
Thus, chaos is not only deterministic, but has a certain "structure"
that can be defined by the form of the entire set of these invariant
curves within the series' domain of convergence.

In our first paper (Efthymiopoulos et al. 2014) we gave one example
of an invariant curve with $c\neq 0$ in a Hamiltonian model.
However, we made no attempt to study such invariant curves in
detail. This is done in the present paper, in which we discuss in
detail the invariant curves $c\neq 0$ in a case of the conservative
2D H\'{e}non map with a hyperbolic point at the origin. Our aims
are: i) to define with precision the domain of series convergence,
ii) to discuss how the series truncation errors scale with
truncation order, iii) to compute the forms of the invariant curves
within this domain, and finally, iv) to study how these forms are
related to particular features of the dynamics close to the
hyperbolic point. Overall, we discuss both the benefits and the
limitations of the method of hyperbolic normal form in
characterizing the local `structure of chaos' around hyperbolic
points.

In section 2 we find a series transformation $\Phi$ of the original
variables $(x,y)$ of the H\'{e}non map around the unstable point
($0,0$) in terms of new canonical variables ($\xi,\eta$), in which
the invariant curves are hyperbolae $\xi\eta =c$. These curves, and
their maps to the original variables, are hereafter called
`invariant Moser curves', since their computation relies on the
method presented in Moser (1956). We find that the transformation is
convergent within the (approximate) limits $|c|\leq$ 0.49 and we
study the accuracy of our calculations by varying the truncation
order of the series and the value of $c$.  We show that, for $c$
close to zero, the invariant Moser curves lie completely outside the
last KAM curve of a central island of stability on the ($x,y$) plane
which exists at some distance from the hyperbolic fixed point.
However, as $c$ approaches the limiting value $0.49$, there are
invariant Moser curves intersecting, or even lying entirely within
the island of stability. In section 3 we study the latter case in
detail, and we explain why this causes no inconsistency with a
different set of invariant curves, namely the KAM curves of the
regular orbits inside the island of stability. Finally, we discuss
how the structure of the invariant Moser curves can be used to
characterize the `stickiness' of the chaotic orbits with initial
conditions near and outside the boundary of the island of stability.
In section 4 we study how the self-intersections of the invariant Moser
curves yield periodic orbits accumulating near the homoclinic points
of the invariant manifold emanating from the central unstable
periodic orbit. We show a numerical example of this accumulation. We
also argue that all the periodic orbits found by such intersections
are regular, i.e. they have been generated by bifurcations from the
stable periodic orbit $S$ at the center of the island of stability.
Section 5 summarizes our conclusions.

\section{Domain of Convergence}
We consider the H\'{e}non symplectic map (H\'{e}non $1969$) 
\footnote{We call it H\'{e}non map because of its similarity with
the original H\'{e}non map (1969), which has $sines$ and $cosines$
instead of hyperbolic $sines$ and $cosines$. Da Silva Ritter et al.
($1987$) used the hyperbolic form of the H\'{e}non map, but they
missed a factor $\sqrt{2}/2$ in their equation (17) that corresponds
to our Eq. (\ref{henmap}). Nevertheless their further equations are
correct.}
\begin{eqnarray}\label{henmap}
x'&=&\cosh(\kappa) x+\sinh(\kappa)y-{\sqrt{2}\over
2}\sinh(\kappa)x^2\nonumber
\\y'&=&\sinh(\kappa) x+\cosh(\kappa)y-{\sqrt{2}\over 2}\cosh(\kappa)x^2
\end{eqnarray}
with $\kappa=1.43$, that was studied by da Silva Ritter et al.
($1987$).
 The origin $(x=y=0)$ is an unstable periodic orbit with
eigenvalues $\lambda_1=e^{\kappa}\approx4.1787$ and
$\lambda_2=e^{-\kappa}=1/\lambda_1\approx0.2393.$ Then we find a
near identity canonical transformation $\Phi=(\Phi_1,\Phi_2)$ of the
form $x=(u+v)/\sqrt{2}$, $y=(u-v)/\sqrt{2}$ with
\begin{eqnarray}\label{phi}
u&=&\Phi_1(\xi,\eta)=\xi+\Phi_{1,2}(\xi,\eta)+...\nonumber\\
v&=&\Phi_2(\xi,\eta)=\eta+\Phi_{2,2}(\xi,\eta)+...
\end{eqnarray}
where $\Phi_{i,s}$ are polynomials of degree \textit{s} in the new
variables $(\xi,\eta)$ such that the mapping (\ref{henmap}) in the
new variables takes the form
\begin{eqnarray}\label{unf}
\xi'&=&\Lambda(c)\xi\nonumber
\\\eta'&=&\frac{1}{\Lambda(c)}\eta
\end{eqnarray}
with
\begin{equation}\label{unfc}
c=\xi\eta=constant
\end{equation}
and
\begin{eqnarray}\label{lam}
\Lambda(c)&=&\lambda_1+w_2c+w_3c^2+...\nonumber\\
\frac{1}{\Lambda(c)}&=&\lambda_2+v_2c+v_3c^2+...
\end{eqnarray}
where $w_s,v_s$ are constants. The details of how to compute the
functions $\Phi$ and $\Lambda$ are given by da Silva Ritter et al.
(1987).

\begin{figure}
\centering
\includegraphics[scale=0.50]{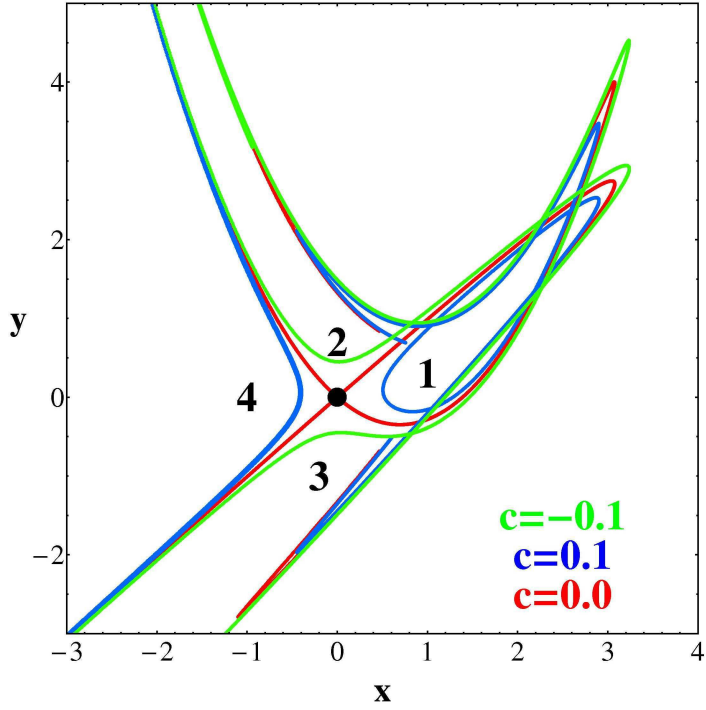}
\caption{The asymptotic curves $c=\xi \eta=0$ (red, mapped in the
original variables ($x,y$)) from the unstable periodic orbit
$(x=y=0)$ and the nearby invariant curves with $c=0.1$ (1 and 4), or
$c=-0.1$ (2 and 3).} \label{fig01}
\end{figure}
The quantity $c=\xi\eta$ is an integral of motion. The cases $c=0$
($\xi=0$~or~$\eta=0$) correspond to the stable and unstable manifold
of the unstable periodic orbit $(0,0)$. When mapped back to the
plane of the original variables $(x,y)$ by the functions $\Phi$ of
Eq.(\ref{phi}), the stable and unstable manifolds are curves
intersecting each other at an infinite number of homoclinic points.
Figure \ref{fig01} shows the form of the stable and unstable
manifolds for the mapping (\ref{henmap}) (red curves). Two
computations are superposed: a purely numerical one, and an
analytical computation using the transformation $\Phi$ truncated at
the order $r=100$, but the two sets of curves are visually
indistinguishable in the scale of Fig.\ref{fig01}. An analysis of
the error of the analytical method is discussed in detail below.
Here, we only note that even low order truncations (e.g. $r=20$, Da
Silva Ritter et al. 1987) allow to find the first few homoclinic
points. Efthymiopoulos et al. (2014) extended their results and
found nine homoclinic points at the truncation order $r=60$. Further
homoclinic points can be found by higher order truncations. Note
that, due to the particular form of the mapping (\ref{henmap}),
intersections of the asymptotic curves exist only on the right of
the origin $(x=y=0)$, while it can be easily shown that the curves
on the left of the origin extend indefinitely to infinity and have
no intersections.

Now, Fig.\ref{fig01} shows also some examples of invariant Moser
curves for $c\neq 0$ (blue and green curves). On the plane
$(\xi,\eta)$, these curves are hyperbolae $c=\xi\eta$. We observe
that the images of such hyperbolae in the variables $(x,y)$, for
small $|c|$, are similar to the corresponding images of the curves
$c=0$. For example, the image of the curve $c=0.1$ (blue curve) on
the right of the origin $(0,0)$ (region 1) follows closely some
parts of the images of both curves $\eta=0,\xi>0$ (unstable
manifold) and $\xi=0,\eta>0$ (stable manifold). Also, the image of
the curve $c=0.1$ intersects itself close to the homoclinic points
of the case $c=0$. On the left of the origin there is another curve
with $c=0.1$ (blue curve, region 4), that follows closely the curves
$\eta=0,\xi<0$ and $\xi=0,\eta<0$. This curve extends indefinitely
to the left, upwards and downwards, without any intersections, in
the same way as there are no homoclinic intersections along the
images of the curves $\eta=0$ and $\xi=0$ on the left of the origin.
Finally, in Fig.\ref{fig01} we plot also parts of the images of two
curves with $c=-0.1$ (green curves), one above the origin $(x=y=0)$
(region 2) and another below the origin (region 3). Both curves
approach the asymptotic curves, coming close to their homoclinic
points.

\begin{figure}
\centering
\includegraphics[scale=0.7]{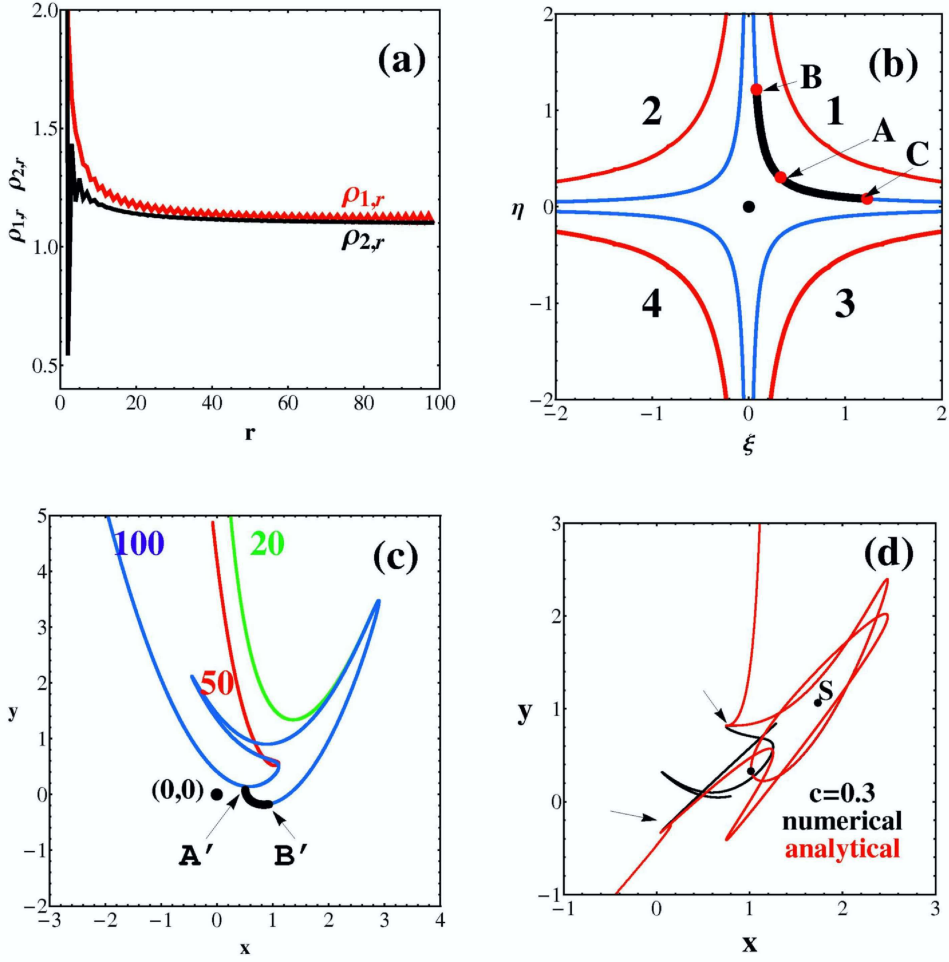}
\caption{(a) The d'Alembert sequences of radii $\rho_{1,r}$ (red)
and $\rho_{2,r}$ (black) for a particular direction with angle
$\phi=3\pi/20$. (b) The limiting hyperbola enclosing the region of
convergence on the $(\xi,\eta)$ plane corresponds to $c\approx\pm
0.49$ (red curves).  The invariant curves $c=\xi\eta=\pm 0.1$ (blue
curves) are well inside the limiting curves. The point $A$ on the
hyperbola $c=0.1$ has coordinates
($\xi_A,\eta_A$=$\sqrt{c},\sqrt{c}$). $B$ and $C$  are its first
images backwards and forward, respectively. (c) A comparison of the
theoretical invariant curves $c=0.1$ by using series truncated at
orders 20 (green), 50 (red) and 100 (blue). The accurate (numerical)
invariant curve coincides with the blue curve. (d) Numerical (black)
versus theoretical (red) invariant curves in the case $c=0.3$. The
theoretical curve using series truncated at order 100 deviates
beyond the points indicated by arrows. } \label{fig02}
\end{figure}

The analytic representation of the invariant curves is successful
within the domain of convergence of the series $\Phi$ and $\Lambda$.
We compute a numerical approximation of the boundary of the domain
of convergence in the same way as in Efthymiopoulos et al. (2014).
Namely, let us consider, for example, the series $\Phi_1$, which has
the form
\begin{equation}
 \Phi_1(\xi,\eta)=\sum_{r=1}^{\infty}\sum_{k=0}^{r}
\Phi_{1,k,r-k}\xi^k\eta^{r-k}~~
\end{equation} with real
coefficients $\Phi_{1,k,r-k}$. Let $\phi$ be the angle of a fixed
direction $\phi=\tan^{-1} (\eta/\xi)$. We write $\xi,\eta$ in polar
coordinates and we define the absolute sums:
\begin{equation}
g_{1,r}(\phi)=\sum_{k=0}^r
\bigg|\Phi_{1,k,r-k}\cos^k\phi\sin^{r-k}\phi\bigg|
\end{equation}
as well as a `d'Alembert sequence of radii':
\begin{equation}\label{dal}
\rho_{1,r}(\phi)=\left({g_{1,r}(\phi)\over
g_{1,r+2}(\phi)}\right)^{1/2}~~.
\end{equation}
Assume that the limit
\begin{equation}\label{rhodal}
R_1(\phi)=\lim_{r\rightarrow\infty}\rho_{1,r}(\phi)
\end{equation}
exists. Then, the series $\Phi_1$ is analytic in the polydisk
$$
D_{1,\phi}=\left\{
|\xi|<R_1(\phi)|\cos(\phi)|, ~~~
|\eta|<R_1(\phi)|\sin(\phi)| \right\}~~.
$$
The complete domain of analyticity of the series $\Phi_1$ around the
origin is given by the union of polydiscs ${\cal D}_1=
\cup_{0\leq\phi<2\pi}D_{1,\phi}$. The intersection of the domain of
analyticity ${\cal D}_1$ with the real plane $(\xi,\eta)$ has a
limiting boundary defined parametrically as a function of $\phi$ via
the equations:
\begin{equation}\label{bdrconv}
\xi=\xi_{1}(\phi) =R_1(\phi)\cos\phi,~~~ \eta=\eta_{1}(\phi)
=R_1(\phi)\sin\phi~~~.
\end{equation}
In the same way we can determine a sequence of d'Alembert radii
$\rho_{2,r}(\phi)$ of the series $\Phi_2$, tending to asymptotic
limiting values $R_2(\phi)$ as $r\rightarrow\infty$. Figure
\ref{fig02}a shows an example of the convergence of the d'Alembert
sequence of radii $\rho_{1,r}$ and $\rho_{2,r}$ for the angle
$\phi=3\pi /20$. We observe that the convergence of both sequences
$\rho_{1,r}$ and $\rho_{2,r}$ towards a limiting value is fast after
approximately the order $r=20$. Furthermore, we find
$\rho_{1,100}=1.12$, $\rho_{2,100}=1.11$, indicating that both
sequences tend to approximately the same limiting value
$R_1(3\pi/20)\simeq R_2(3\pi/20)$. In fact, we may use a numerical
extrapolation technique to better estimate $R_1$ or $R_2$. Namely,
denote by
\begin{equation}\label{seqerr}
\delta\rho_{1,r} = \rho_{1,r}-R_1
\end{equation}
the sequence of differences between the $r-th$ value of the
d'Alembert sequence of radii $\rho_{1,r}$ and the limiting value
$R_1$. Using as an initial estimate $R_1\simeq \rho_{1,100}$, we
compute the differences $\Delta\rho_{1,r}=\rho_{1,r}-\rho_{1,100}$
for $r<<100$ (we set $r<50$ in the computation). Then, we
numerically observe that, apart from a first few transient orders
$r$, the sequence $\Delta\rho_{1,r}$ scales approximately with $r$,
up to $r=50$, as a power-law, namely one finds that
$|\Delta\rho_{1,r}|\simeq \Delta_1 r^{-q}$, where the constants
$\Delta_1,q>0$ can be found by best-fitting. Assuming, now, that the
estimate $\delta\rho_{1,r}\approx\Delta_1 r^{-q}$ extends to all the
true errors $\delta\rho_{1,r}$ for $r$ up to the maximum one, the
sequence
$$
\rho'_{1,r}=\rho_{1,r}-\Delta_1 r^{-q}~~
$$
should be constant for large $r$. We then check numerically that
$\rho'_{1,r}$ is indeed constant to about three significant figures
beyond $r>20$. This allows to better estimate $R_1$ as $R_1\simeq
\rho'_{1,100}$. Even more precise estimates can be obtained by
iterating the above extrapolation algorithm. In the case of the data
of Fig.\ref{fig02}a, implementing the same algorithm to both
sequences $\rho_{1,r}$ and $\rho_{2,r}$, we find
$\rho'_{1,100}=1.09554$, $\rho'_{2,100}=1.09555$, indicating that
$R_1=R_2=R=1.095$ with a precision of three significant digits.

The limiting value of $R(\phi)$ defines the limiting point of
Eq.(\ref{bdrconv}). Repeating the computation for many different
values of $\phi$ in the interval $0\leq\phi<2\pi$ allows to obtain a
numerical approximation of the boundary of the domain of
convergence. This is shown with a red curve in Fig.\ref{fig02}b. The
boundary is found to be hyperbola-like. In fact, computing $c(\phi)
= R^2(\phi)\cos(2\phi)/2$ with our numerical estimates of $R(\phi)$
based on the extrapolation algorithm, we find that $c(\phi)$ is {\it
remarkably constant}, namely $c(\phi)\simeq 0.49$ with a variation
of less than $3\times 10^{-3}$ over the whole range of values of
$\phi$. Note that, as shown in da Silva Ritter et al. (1987), one
has $R\rightarrow\infty$ in both limits $\phi\rightarrow 0$ or
$\phi\rightarrow\pi/2$, i.e., the domain of convergence extends to
infinity along the $\xi$ and $\eta$ axes representing the invariant
manifolds. For values of $\phi$ very close to the limiting ones, $0$
and $\pi/2$, the convergence of the d'Alembert sequences of radii is
extremely slow. Even so, using the extrapolation algorithm yields
the value $c=0.492$ even for $\phi=\pi/2- 1.57\times 10^{-5}$. In
conclusion, after correcting with the extrapolation algorithm, our
numerical evidence is that the boundary of the domain of convergence
is very close to a hyperbola with $c=c_{max}\simeq 0.49$ for all
values of $\phi$. The same is true for the convergence of the series
$\Lambda(c)$ (Eq. (\ref{lam})), where we have found, using
d'Alembert sequences, that the boundary of the domain of convergence
corresponds again to $c=c_{max}\simeq 0.49$.

Since the series $\Phi$ are convergent for all $(\xi,\eta)$ with
$|\xi\eta|<c_{max}$, we may in principle use them to find the image,
under $\Phi$, of all hyperbolae $\xi\eta=c$ with $|c|<c_{max}$, in
the plane $(x,y)$. However, in computing these images there arise
practical limitations due to the errors by the use of only finite
truncations of the series. A control of the error can be achieved by
a combination of analytical and numerical propagation of the orbits,
as exemplified in Figs.\ref{fig02}b,c. In Fig.\ref{fig02}c, the arc
$A'B'$ is the image under $\Phi$ of an arc $AB$, shown in
Fig.\ref{fig02}b, which lies on the hyperbola $c=0.1$. The point $A$
is chosen in the diagonal, i.e. $\xi_A=\eta_A=\sqrt{0.1}$, while the
point $B$ is the first backward image of the point $A$ under the
normal form equations (\ref{unf}). In Figs.\ref{fig02}b,c, both the
point $B$ as well as the image $A'B'$ of $AB$ were computed by
taking the truncation of the series (\ref{unf}), (\ref{lam})  and
(\ref{phi}), at the order $r=100$. We can have an estimate of the
truncation error by comparing the images of the arcs $A'B'$ when
computed at different truncation orders.  We find that the
differences between the images of the points of the arc $A'B'$ at
the orders $r=20$ and $r=60$ are about $10^{-9}$ and between $r=60$
and $r=100$ are even smaller. Therefore, in the case $c=0.1$ the
accuracy of the computed arcs $A'B'$ truncated at order $100$ is
sufficient for further calculations.

We hereafter call the arcs $AB$ (and their images $A'B'$), `Moser
arcs', namely they are arcs belonging to an invariant Moser curve.
Their main property is the following: if we iterate the Moser arc
$AB$ backwards using the inverse of the normal form equations
(\ref{unf}), we obtain the whole upper branch of the invariant Moser
curve $c=0.1$ from the point $A$ upwards. In the same way, if we iterate
the arc $A'B'$ backwards using the inverse of the original H\'{e}non
mapping (Eqs.(\ref{henmap})), we obtain the image, under $\Phi$, of
the upper branch of the Moser curve $c=0.1$ in the plane $(x,y)$.
In the same way, we may obtain the image, under $\Phi$, of the lower
branch of the invariant Moser curve $c=0.1$ from the point A downwards.
Namely, we compute the Moser arc $AC$, where $C$ is the first
forward image of $A$ under the Eqs.(\ref{unf}). Then, we use
the series $\Phi$ at a high truncation order to compute the image
$A'C'$ with accuracy, and then we propagate the arc $A'C'$
forward using the original mapping equations (\ref{henmap}).

The above process allows to find numerically the images, in the
plane $(x,y)$ under $\Phi$, of the whole Moser invariant hyperbolae
for $|c|<c_{max}$. We call these curves `numerical' or `exact
invariant Moser curves' (i.e. they are nearly free of series
truncation errors). We notice that the `numerical' or `exact' curves
cannot be defined independently of the mapping $\Phi$. The reason is
that although the H\'{e}non map (\ref{henmap}) is exact and gives
the iterate $(x',y')$ of any given point $(x,y)$, only the rules of
Eqs.(\ref{phi}), (\ref{lam}) and (\ref{unf}) allow to define an
\textit{invariant curve} (i.e. a curve mapped onto itself) joining
the points $A'$ and $B'$ on the $(x,y)$ plane. Furthermore, the
whole computation relies on that the initial invariant arc $A'B'$
should be possible to compute with precision via the arc $AB$ using
exclusively a high order truncation of the series $\Phi$.

On the other hand, we can also compute purely analytical images of
the invariant Moser curves by implementing the (truncated) series
$\Phi$ directly to different points on the hyperbolae {\it also
outside} the arcs $CAB$ (Fig.2b), i.e. without any numerical
propagation. A comparison of the obtained images, for $c=0.1$, at
the truncation orders $r=20$, $r=50$, and $r=100$ is shown in
Fig.\ref{fig02}c. We see that, using exclusively the series $\Phi$
truncated at various orders, we find only parts of the `exact'
invariant curve, while beyond that part the theoretical curve
deviates considerably. E.g. if we use series truncated at order
$r=20$, the deviation appears after the first oscillation (green
curve). At the truncation order $r=50$ the deviation starts beyond
the second oscillation (red curve). Finally, if we use series
truncated at order $r=100$ we follow the exact curve well beyond the
limit of applicability of the red curve, giving now also the third
oscillation of the exact curve (blue curve).

Similar results are found for the initial arc $A'C'$, which is the
image under $\Phi$ of the arc $AC$ of Fig.\ref{fig02}b.

The accuracy decreases for larger (absolute) values of $c$. For
example, in Fig.\ref{fig02}d we compare the analytical invariant
curve truncated at order 100 (red) for $c=0.3$ with the exact
numerical curve (black) and we find large deviations after the
second oscillation. For even larger $c$, close to the limit
$c=0.49$, the theoretical curve deviates even before the first
oscillation of the exact curve.

\begin{figure}
\centering
\includegraphics[scale=0.7]{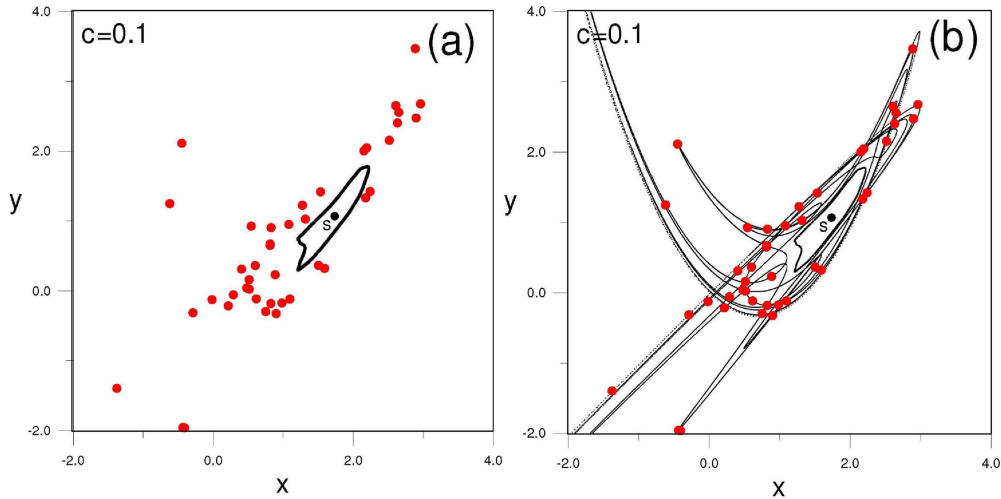}
\caption{(a) The successive iterations of two orbits having initial
points on the invariant curve $c=0.1$ seem to be distributed
randomly around the last KAM curve of a central island of
stability (b) The scattered points of Fig.\ref{fig11}a  belong in
fact to the invariant curve $c=0.1$. } \label{fig11}
\end{figure}
If we compute the length $s$ along an exact curve (around the point
$A'$) that is accurately represented by the analytical computation
of the series up to the truncation order $r$, we typically find
$s\sim\log(r)$ (see also Efthymiopoulos et al. 2014). This estimate
poses the overall limitations of the whole method of analytic
(normal form) computations. Namely, the number of terms required to
compute a part of fixed length $s$ along an invariant Moser curve
grows {\it exponentially} with $s$. This is consistent with the fact
that these curves represent mostly chaotic motions. Namely, one
requires an exponentially growing amount of information (in the
form, here, of series coefficients) in order to compute analytically
larger and larger lengths of chaotic orbits. For example,
Fig.\ref{fig11}a shows some scattered points, which are successive
iterates of two chaotic orbits with initial conditions on the
invariant Moser curve $c=0.1$. The thick curve represents the last
KAM curve around an island of stability which surrounds a stable
periodic orbit S at a certain distance from the origin.  This last
KAM curve plays a particular role, since it constitutes an absolute
barrier separating orbits which lie entirely in the interior or the
exterior of the curve. In particular, no chaotic diffusion leading
to a crossing of the last KAM curve can occur. On the other hand,
there are invariant Moser curves which intersect the last KAM curve.
Thus, different initial conditions along one such invariant Moser
curve lead in general to different orbits, which lie either entirely
inside or entirely outside the last KAM curve.  The connection
between such invariant Moser curves and  the behavior of orbits
inside the island of stability will be discussed in section 3 below.
The orbits of Fig.\ref{fig11}a, instead, lie entirely outside the
last KAM curve. A measurement of their Lyapunov characteristic
exponents shows that  these orbits are chaotic. The fact that the
Lyapunov characteristic numbers are positive implies that successive
iterates of points are far away from each other and appear as
randomly scattered in a large neighborhood of the periodic orbit
($x=y=0$). However, in Fig.\ref{fig11}b we see that all the
scattered points belong to the same invariant Moser curve $c=0.1$.
Thus, in principle, these points can be found not only numerically
but also analytically with sufficient accuracy by using the series
$\Phi$ at a high truncation order. Nevertheless, one needs to
compute a ${\cal O}(\exp N)$ number of terms in order to compute
accurately the $N-$th iterate of each orbit. Thus, the orbits become
practically non-computable beyond some iterations by using
exclusively any reasonable order truncations of the series
(\ref{phi}), (\ref{lam}) and (\ref{unf}).

Even so, the fact that these iterates are arranged
along an invariant curve allows to define a `structure of chaos',
given by the ensemble of all the invariant Moser curves laying in
the chaotic domain around the origin. This structure is important
because it determines the dynamics of recurrences in the chaotic
domain for all the orbits which do not belong to the asymptotic
curves of the unstable periodic orbit at the origin (cf. Contopoulos
and Harsoula 2008).

\begin{figure}
\centering
\includegraphics[scale=0.4]{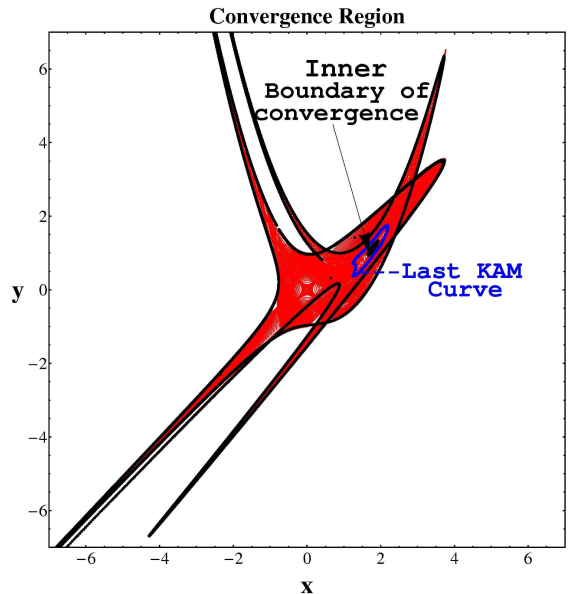}
\caption{The region of convergence in the plane $(x,y)$
corresponding to the image, under $\Phi$, of the region inside the
limiting curves of Fig. 2b. The upper right limiting curve (1) of
Fig. 2b is mapped into the island of stability, while the curves
(2), (3), (4) are mapped into  curves that surround the outer limits
of the region of convergence. E.g. the line (2) is mapped into a
curve that makes an infinite number of oscillations on the lower and
upper side (most of them are close to the asymptotic curves from the
periodic orbit $(0,0)$ on the left (upwards and downwards) and are
not seen separately in this figure). A similar form is taken by the
image of the curve (3), while the curve (4) is like a hyperbola
without oscillations. The inner boundary of the region of
convergence is inside the last KAM curve.} \label{fig04}
\end{figure}
Using, now, the above method of `exact' curves, in Fig.\ref{fig04}
we plot the image, under $\Phi$, of the domain of convergence in the
interior of the limiting boundary curves of Fig.\ref{fig02}b. The
corresponding region of convergence in the plane $(x,y)$ is shown in
red. The plot is obtained by computing the images of a foliation of
hyperbolae for $-c_{max}<c<c_{max}$. The main remark is the
following: the blue closed curve in the same plot is, again, the
last KAM curve of the island of stability around the stable periodic
orbit S at $x\approx 1.736$, $y\approx 1.065$. We observe that,
besides the chaotic domain around the origin, the region of
convergence of the hyperbolic normal form series extends even {\it
inside a part of the island of stability}. In particular, we find
that there are invariant Moser curves intersecting, or even lying
{\it entirely} within the interior of the last KAM curve.  Let us
note that by computing `numerical' Moser curves for $c$ very close
to $c=0.49$ we find that these curves leave a small hole around the
central elliptic fixed point, i.e., the domain of convergence does
not reach the center of the central island of stability. Even so,
the invariant Moser curves should be able to characterize not only
chaotic motions, but also some features of the regular dynamics
close to or inside the main island of stability. To this question we
now turn our attention.

\section{Island of stability - stickiness}

In all computations of this section, the following conventions hold:

i) We refer to a `last KAM curve' which is the same as the one shown
in Fig.\ref{fig11}a. This curve was computed by a detailed
calculation of `rotation numbers' and `twist numbers', using a
method developed by Voglis and Efthymiopoulos (1998).

ii) All invariant Moser curves were computed by the numerical method
of the previous section, namely iterations of Moser arcs
$C'A'B'=\Phi(CAB)$ defined on the corresponding hyperbolae labeled
by different values of $c$.

 Since the last KAM curve is an absolute barrier separating its
interior from the exterior under the mapping (\ref{henmap}), it
follows immediately that three cases can be distinguished: a) an
initial Moser arc $C'A'B'$ lies entirely outside the last KAM curve.
Then, every initial condition on this arc leads to an orbit lying
entirely outside the last KAM curve. Thus, the {\it whole}
corresponding invariant Moser curve lies entirely outside the last
KAM curve. b) Similarly, initial Moser arcs lying entirely inside
the last KAM curve define invariant Moser curves lying entirely
inside the last KAM curve. However, if c) an initial Moser arc lies
partly inside and partly outside the last KAM curve, then, by
implementing the method described above, all initial points on the
arc which are inside or outside the last KAM curve define two
disjoint parts of the corresponding invariant Moser curve, which are
inside and outside the last KAM curve respectively. Such curves play
a role in the characterization of the phenomenon of `stickiness', as
will become clear by specific examples below.

\begin{figure}
\centering
\includegraphics[scale=0.3]{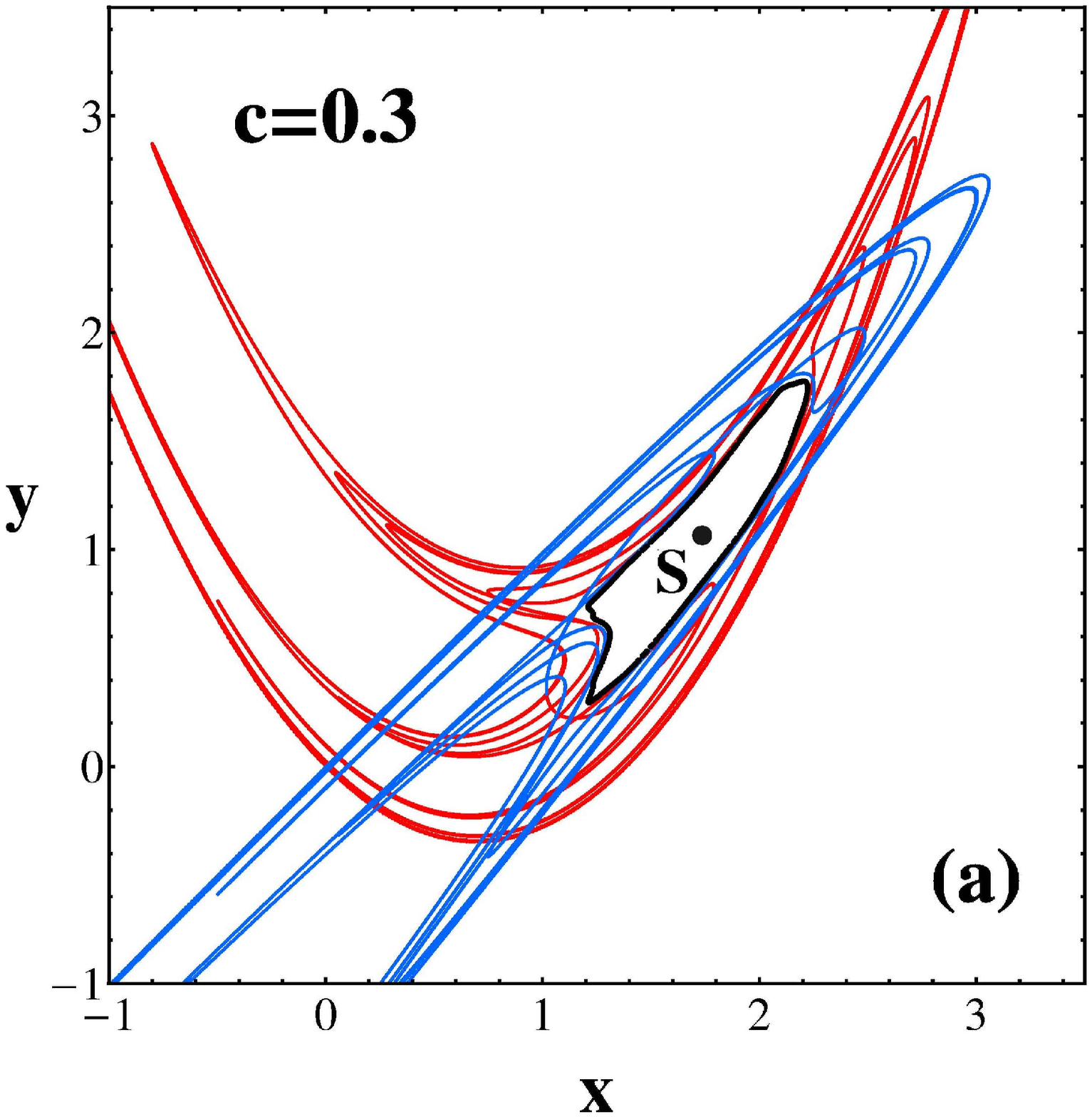}
\includegraphics[scale=0.3]{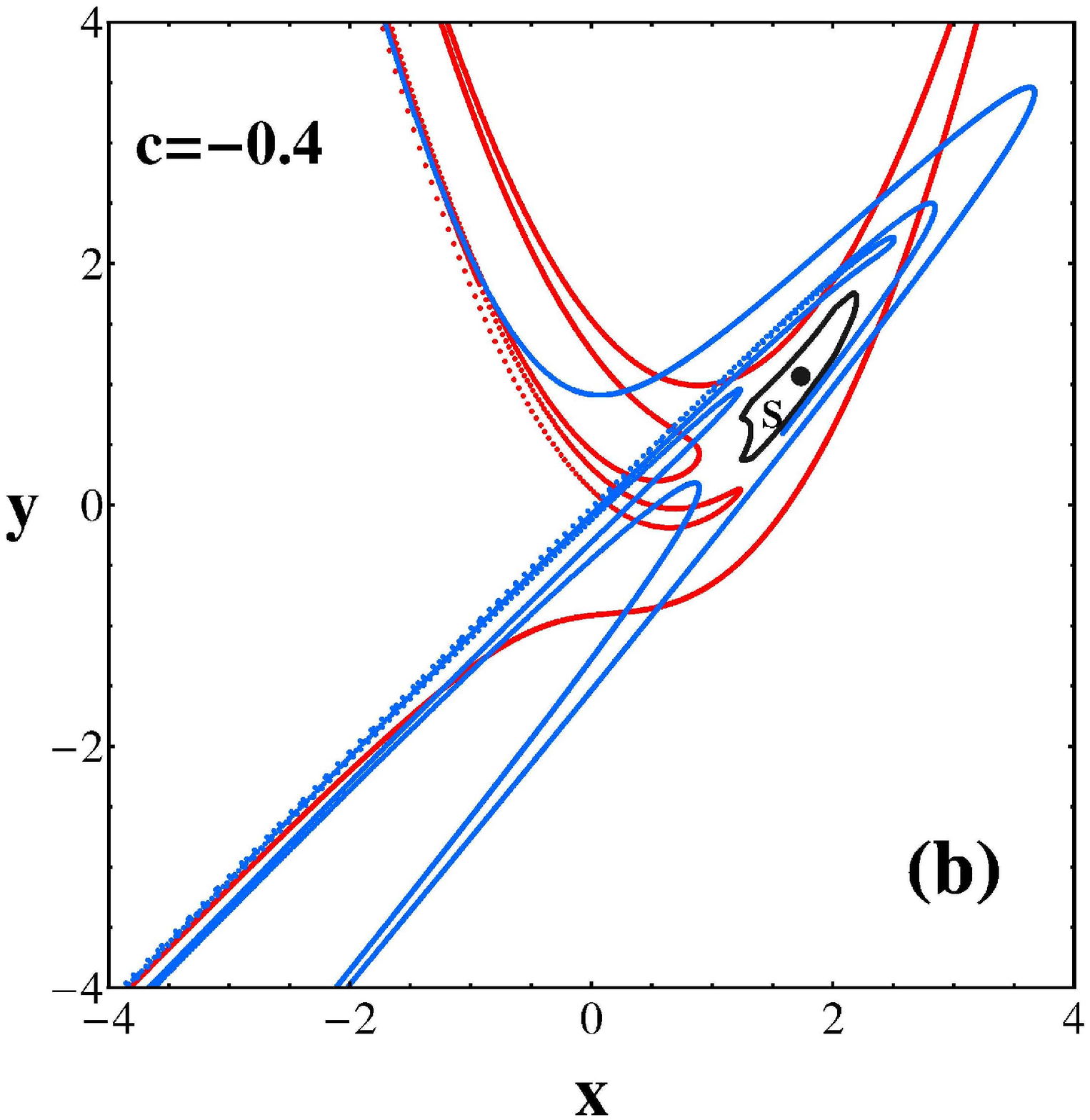}
\caption{(a) The invariant curve $c=0.3$ fills most of the
convergence region of Fig.\ref{fig04}, outside the island of
stability defined by the last KAM curve around the stable point $S$.
(b) Two invariant curves with $c=-0.4$ (blue and red) also fill most
of the convergence region outside the island of stability $S$.}
\label{fig07}
\end{figure}

As a first example, the numerical Moser curve $c=0.3$ fills most of
the region of convergence of Fig.\ref{fig04} (Fig.\ref{fig07}a). The
backward images of the initial arc $A'B'$ are given in red and the
forward images of $A'C'$ are given in blue. The higher arcs of the
curve $c=0.3$ approach the last KAM curve (black curve) around the
stable periodic point S. However as the initial arc $C'A'B'$  lies
outside the last KAM curve, its images cannot intersect the last KAM
curve. This means that the whole invariant Moser curve $c=0.3$ is
outside the last KAM curve.
\footnote{Note the distinction of terminology: both The `KAM curves'
and the `Moser curves' are invariant, i.e., they are mapped onto
themselves under the mapping (\ref{henmap}). However, as discussed
in the text, their properties are different.}

Similar invariant curves are formed when $c<0$. In Fig.\ref{fig07}b
we plot the case $c=-0.4$. This curve also does not enter into the
island $S$ because its initial arc $C'A'B'$ is far outside the last
KAM curve.

\begin{figure}
\centering
\includegraphics[scale=0.3]{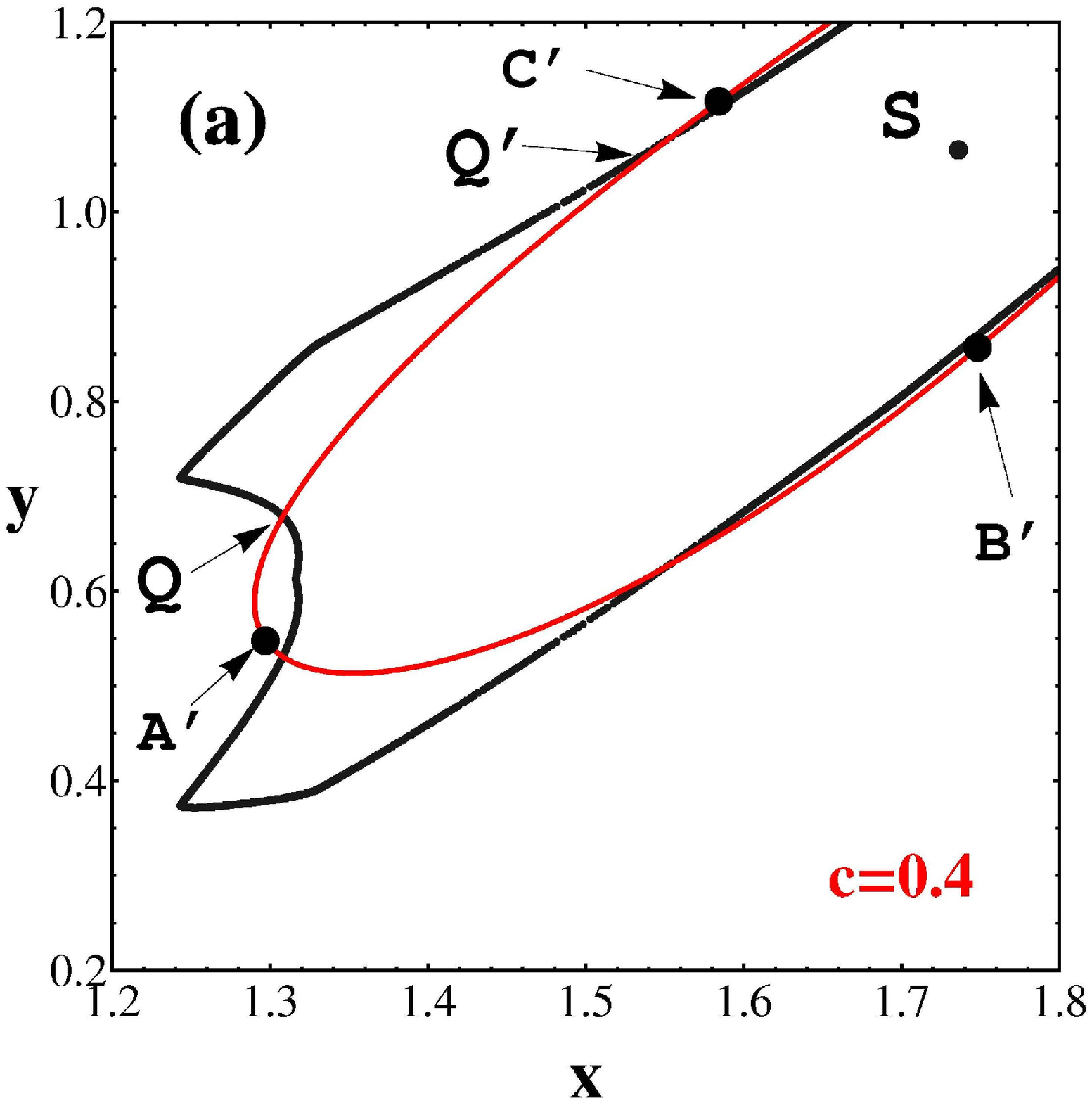}
\includegraphics[scale=0.3]{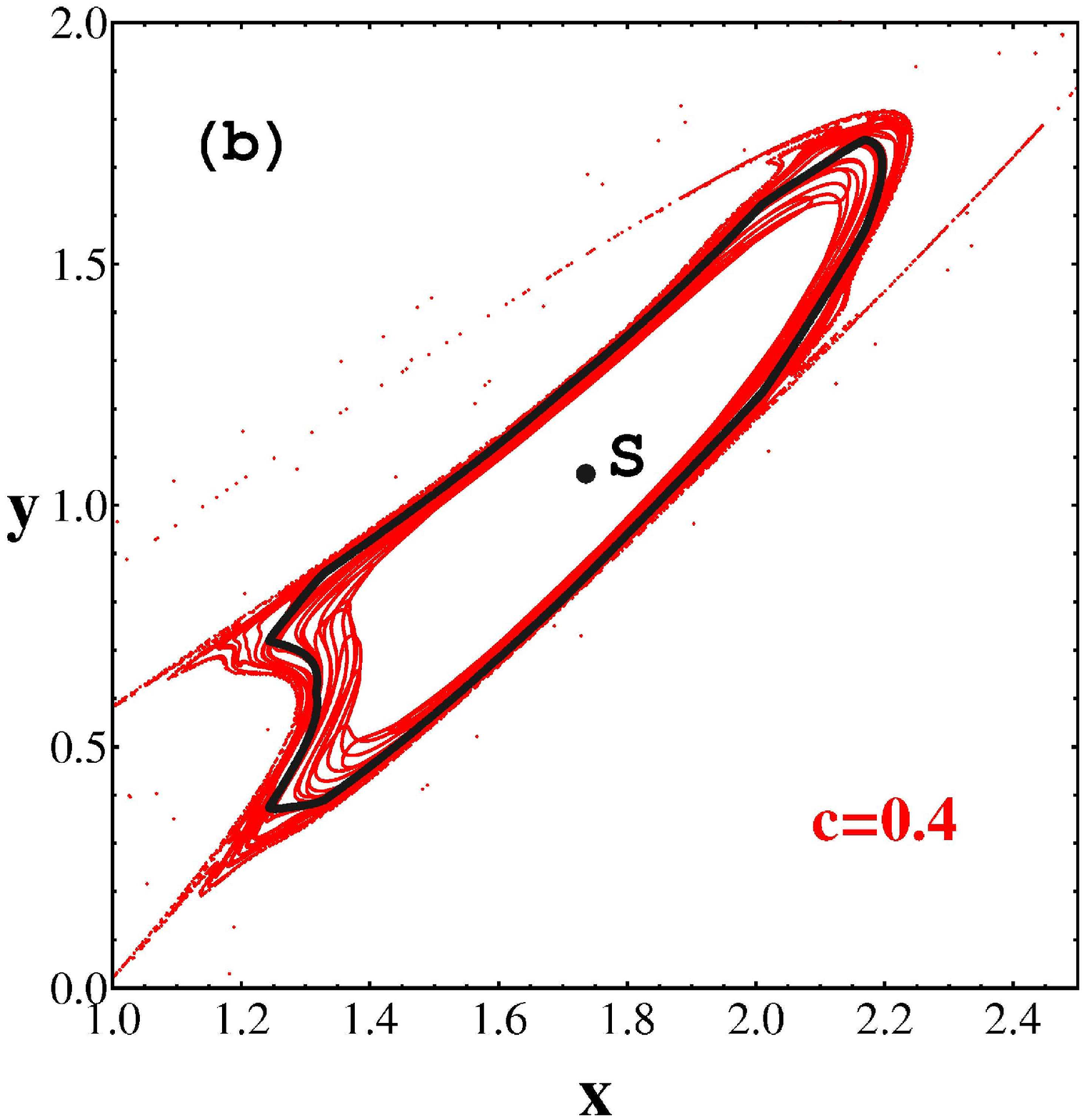}
\caption{The invariant curve with curve $c=0.4$ (a) The initial arc
$C'A'B'$ of this curve intersects the last KAM curve. We note $Q$
and $Q'$ the points of intersection of the arc $A'C'$ with the last
KAM curve. (b) Further numerical images (using map (\ref{henmap}))
of the initial arc $C'A'B'$ of this curve extend to large distances
outside the last KAM curve. Inside the last KAM curve they fill a
region which is limited by another KAM curve.} \label{fig08}
\end{figure}
On the other hand, the invariant curve $c=0.4$ does intersect the
last KAM curve around S (Figs. \ref{fig08}ab), because the original
arc $C'A'B'$ of this curve intersects the last KAM curve. In fact in
Fig.\ref{fig08}a we see $4$ points of intersection. We note $Q$ and
$Q'$ the points of intersection of the arc $A'C'$ with the last KAM
curve. The images of these points, infinite in number, are also
points of intersection.

If we calculate numerically many iterations of the initial arc
$C'A'B'$ of the curve $c=0.4$ we find that part of it is inside the
last KAM but it has also extensions outwards and probably it covers
most of the region of convergence of Fig.\ref{fig04}
(Fig.\ref{fig08}b). On the other hand the curve $c=0.4$ enters also
into the island of stability, therefore it intersects an infinity of
closed KAM curves around S, inside the last KAM curve. Every point
of the invariant curve $c=0.4$ inside the last KAM curve either
belongs to a KAM curve or is between two KAM curves.
\footnote{In fact from the theory of the KAM curves (see, e.g.,
Contopoulos 2002) it is known that although the set of KAM curves
has a positive measure there are infinitely many points between any
two KAM curves not belonging to a KAM curve.}
The images of those points are also on KAM curves or between two KAM
curves. Thus the initial arc $C'A'B'$ of the invariant curve inside
the last KAM curve has infinite images forming a ring around the
point $S$.

 Every KAM curve can be found
exactly by means of analytic expressions around the central point
$S$ (KAM theorem). However the present theory of the Moser invariant
curves (around the unstable point (0,0)) is valid also between the
KAM curves, where no expressions  for KAM curves exist. The domain
filled by the curve $c=0.4$ is a ring whose inner boundary is a
limiting KAM curve (inside the last KAM curve, Fig.\ref{fig08}b).
\footnote{It should be emphasized that there is no inconsistency
between the fact that orbits with given initial conditions can be
represented by two different types of convergent series. In fact,
the hyperbolic normal from series converge in open domains of
initial conditions, but their convergence is non-uniform along any
particular KAM orbit. On the other hand the series of the KAM theory
are convergent only on a Cantor set of initial conditions, even
while their convergence along any particular KAM orbit is uniform. }
The transition  between initial arcs $C'A'B'$ of various $"c"$ that
are completely outside the last KAM curve and initial arcs that
intersect the last KAM curve occurs close to $c=0.32.$

\begin{figure}
\centering
\includegraphics[scale=0.3]{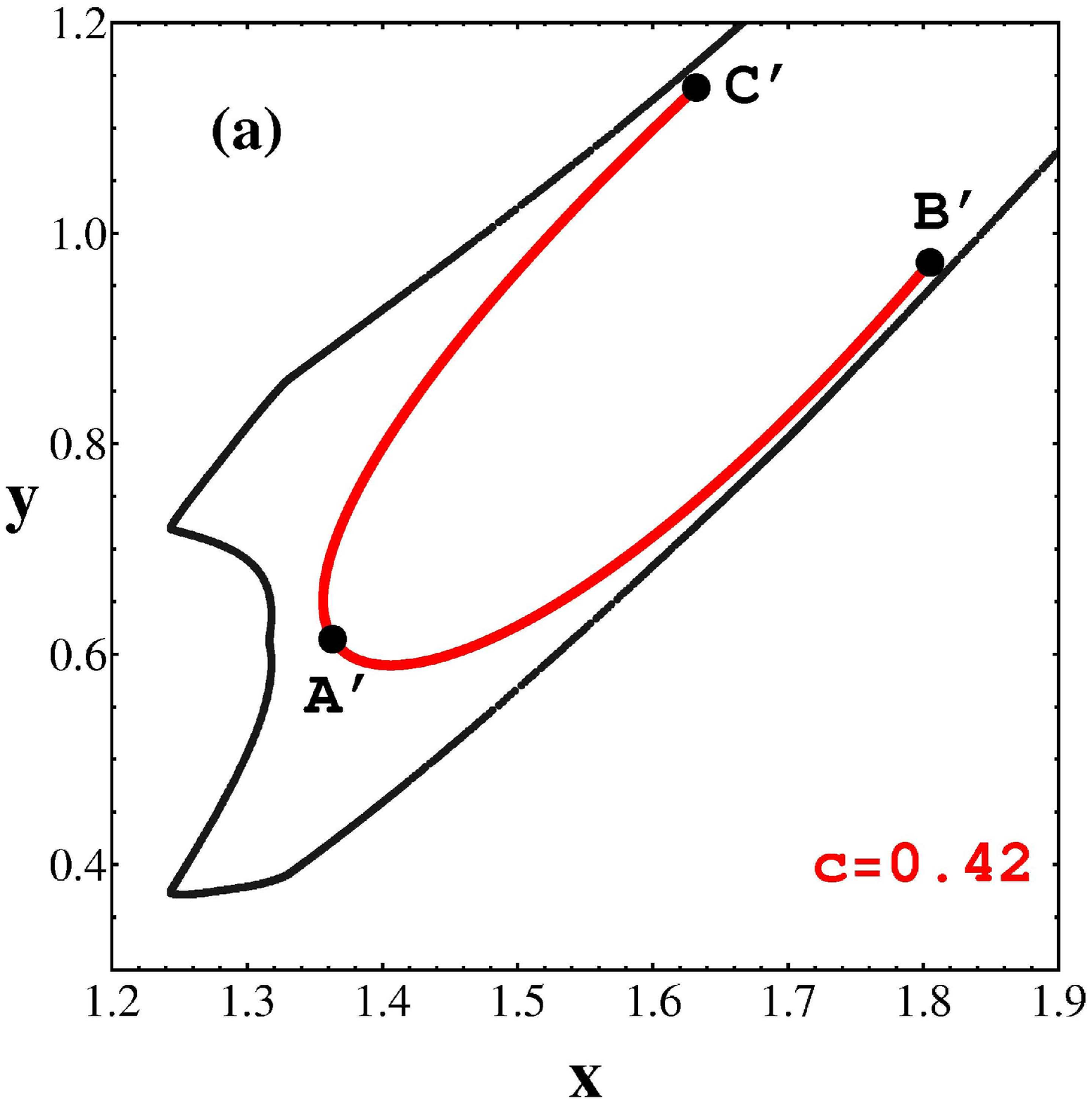}
\includegraphics[scale=0.3]{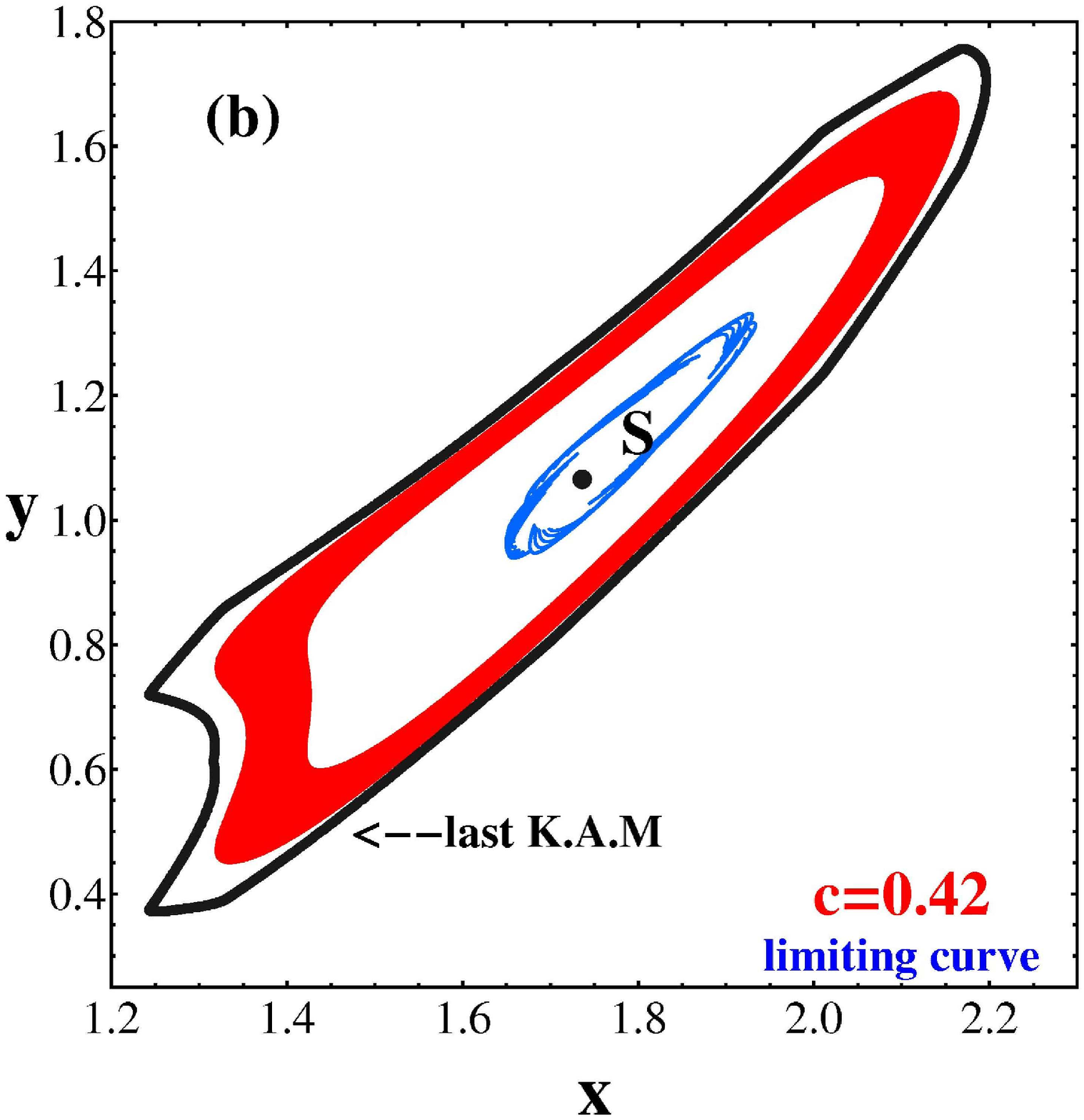}
\caption{(a) The initial arc $C'A'B'$ of the curve $c=0.42$ is
marginally inside the last KAM curve (b) The curve $c=0.42$ fills a
ring (red) completely inside the last KAM curve. The inner boundary
of the region of convergence (blue) is very close to the invariant
curve corresponding to the hyperbola with $c=0.49$.} \label{fig09}
\end{figure}
If we increase \textit{c}~beyond $c=0.4$ we find cases in which the
initial arc $C'A'B'$ is completely inside the last KAM curve. Figure
\ref{fig09}a is such a case for $c=0.42$. Any other KAM curve
passing through points of this initial arc has also infinitely many
points of intersection with the images of the invariant Moser curve
$c=0.42$. Finally, the ring formed by this curve (red) if we take a
large number of iterations of the initial arc $C'A'B'$, is limited
on the outside and on the inside by KAM curves. In Fig.\ref{fig09}b
we also plot, for comparison, the region filled by the inner
boundary of the convergence region (blue) which corresponds to
$c\approx0.49$.

It was shown above that the invariant Moser curves with $c$
sufficiently close to the limiting value $c_{max}=0.49$ lie entirely
within the main island of stability. As we decrease $c$, we find a
critical value $c_{crit}$ such that the invariant Moser curve with
$c=c_{crit}$ comes tangent with the last KAM curve surrounding the
main island of stability. We find $c_{crit}\simeq 0.41$ (compare
Figs.\ref{fig08}a for $c$=0.4 and \ref{fig09}a for $c$=0.42).

We now show how the properties of the invariant Moser curves in the
neighborhood of the critical curve $c=c_{crit}$ allow to characterize
the {\it stickiness} of chaotic orbits with initial conditions
outside, but close to the last KAM curve.

Consider two neighboring invariant curves with
$c_1=c_{crit}+\epsilon$, $c_2=c_{crit}-\epsilon$, for arbitrarily
small $\epsilon$ (positive). Since $c_1>c_{crit}$, the invariant
curve $c=c_1$ lies entirely within the main island of stability. On
the other hand, the endpoint of the initial arc $A'C'$ along the
curve $c=c_2$ is located outside the last KAM curve. Let $Q,Q'$ be
the points where the initial arc along $c=c_2$ intersects the last
KAM curve (see for example Fig.\ref{fig08}a, for $c=0.4$). Then,
along the segments $QA'$ and $Q'C'$ there are some initial
conditions whose images, after a (possibly large) number of
iterations $N$ under the mapping (\ref{henmap}) can be far (at order
unity distance) from the last KAM curve. We will now show that the
above argument implies that the transformation $\Phi$ has the
following property: it is possible to find pairs of neighboring
points $(\xi_1,\eta_1)$ and $(\xi_2,\eta_2)$, separated by an
arbitrarily small distance $\delta$ in the plane $(\xi,\eta)$, whose
images under $\Phi$ are points $(x_1,y_1)$ and $(x_2,y_2)$ separated
by an order unity distance.

\begin{figure}
\centering
\includegraphics[scale=0.35]{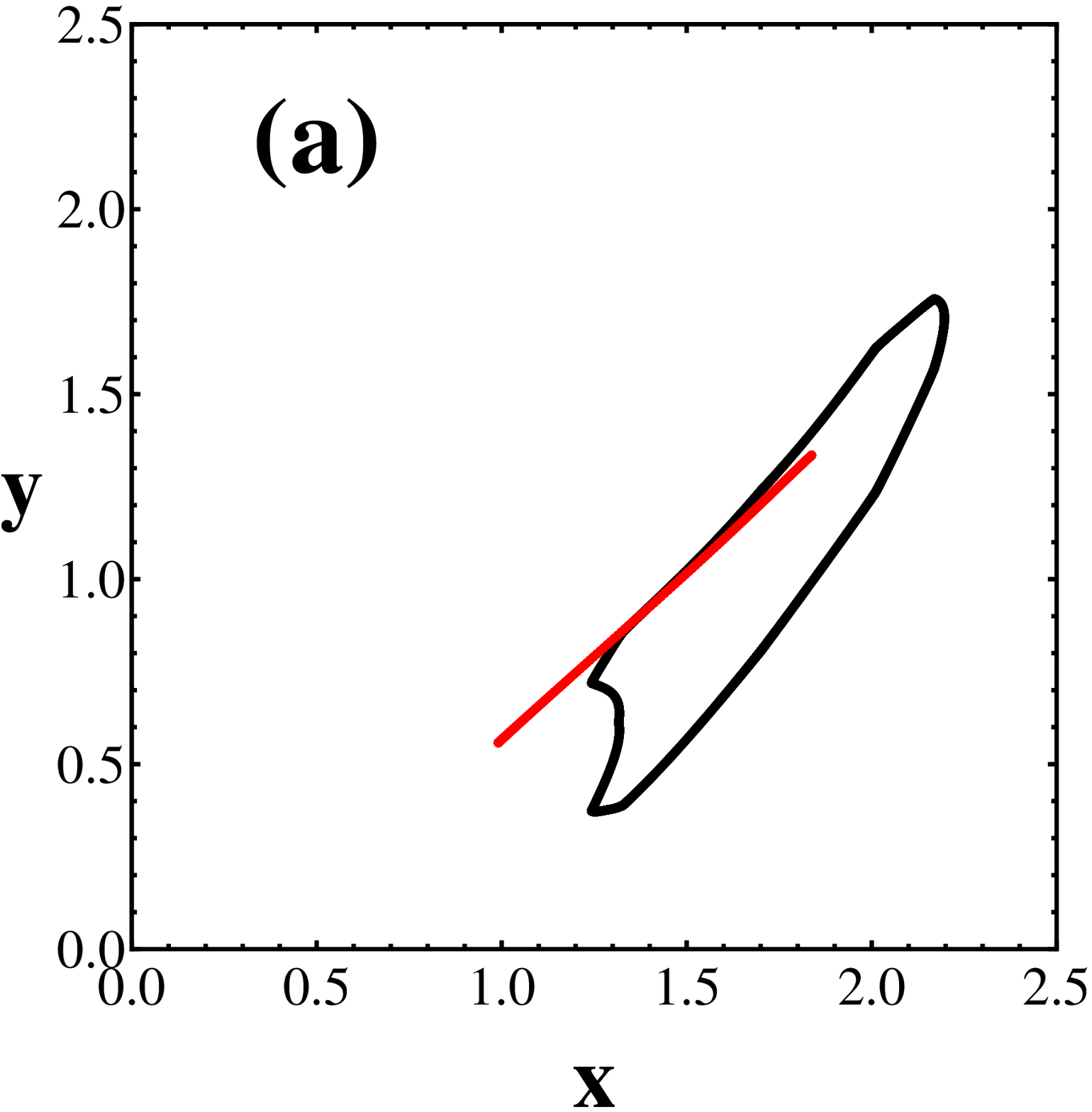}
\includegraphics[scale=0.33]{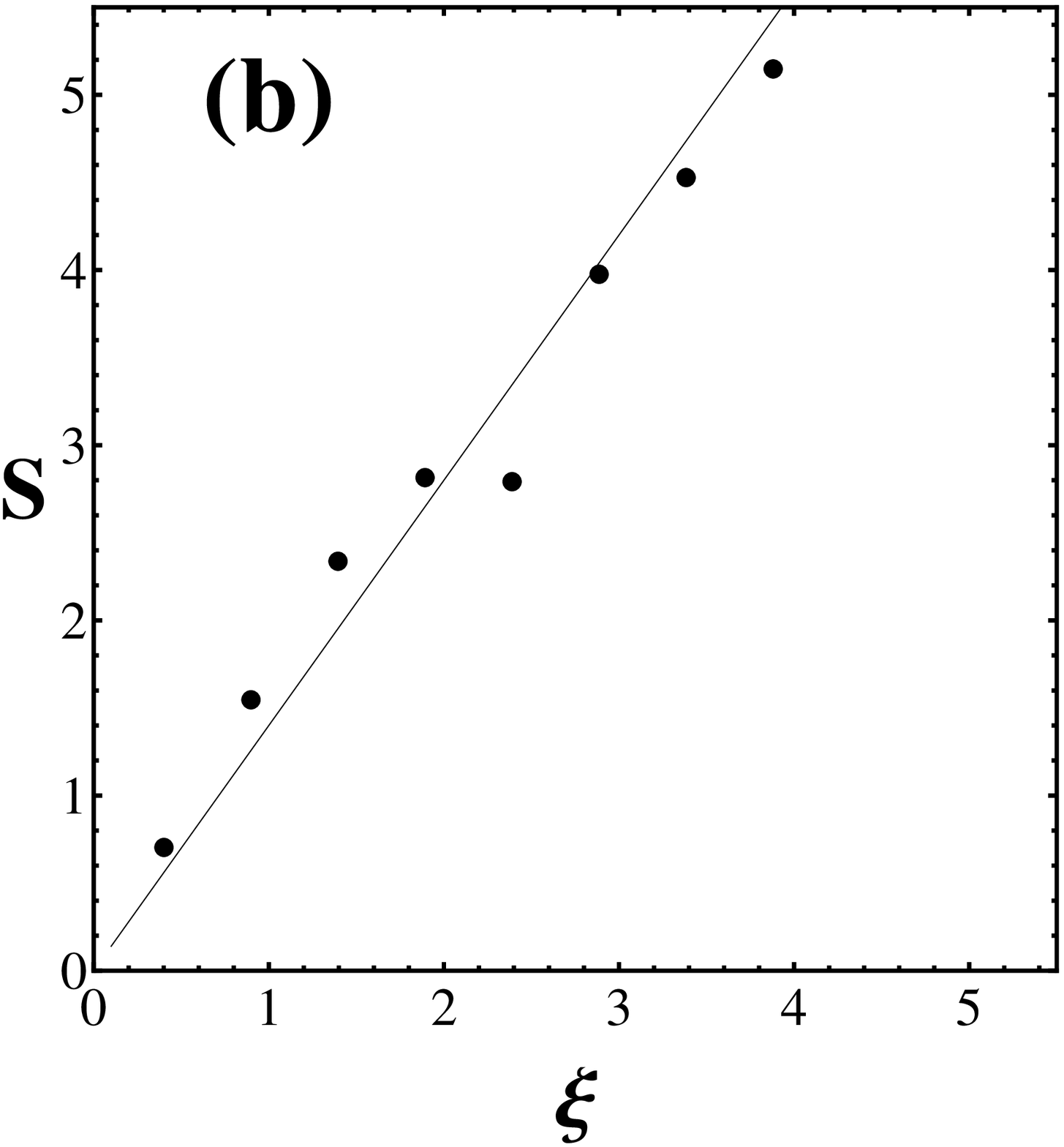}
\includegraphics[scale=0.35]{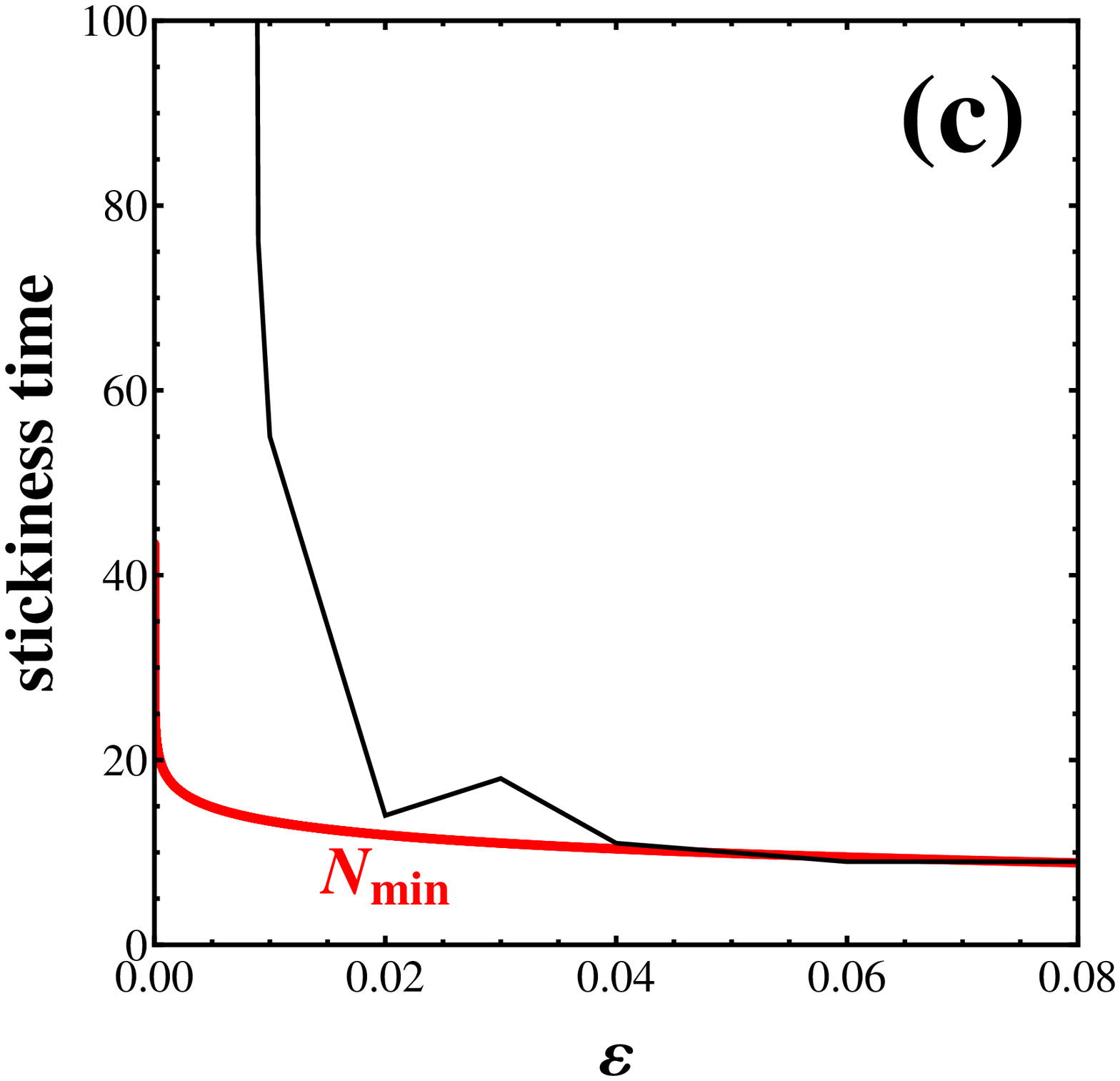}
\caption{ (a) The last KAM curve (black curve) together with the
image, under $\Phi$, of a disc of size $2.6\times10^{-6}$ centered
on the hyperbola with $c=c_{crit}=0.41$ on the ($\xi,~\eta$) plane
(red curve). This is a very elongated ellipse of size about 1.1,
i.e. of order unity. (b) The function $S(\xi_0)$ of Eq. (11) is a
power law of the form $S(\xi_0)\sim 3.9\xi_0^{1.4}$. (c) The
prediction of the minimum stickiness time (red)) against the true
stickiness time (black) for initial conditions along invariant Moser
curves with $c=c_{crit} -\epsilon$ as a function of $\epsilon$.}
\label{fig19}
\end{figure}
Indeed, consider a point ($\xi_{crit},\eta_{crit}$) on the critical
hyperbola with $c=c_{crit}$, as well as a disc $\Delta$ centered on
the  point ($\xi_{crit},\eta_{crit}$) with radius
$\delta=\eta_1-\eta_{crit}$, where the point ($\eta_1,\xi_{crit}$)
with $\eta_1=c_1/\xi_{crit}$ belongs to the hyperbola with $c=c_1$.
Then, $\delta=(c_1-c_{crit})/\xi_{crit}$=$\epsilon/\xi_{crit}$. The
points with constant $\xi=\xi_{crit}$ and $\eta$ between
$\eta_{crit}$ and $\eta_1$ belong to hyperbolae with $c>c_{crit}$.
  Hence, they are mapped, under $\Phi$,
to invariant Moser curves lying entirely within the last KAM curve.
On the other hand, it is possible to find points with arbitrarily
large $\xi_2$= $\xi_{crit}$, i.e., arbitrarily small $\delta$, such
that $(x_2,y_2)=\Phi(\xi_2=\xi_{crit}, \eta_2=c_2/\xi_2)$ lies at an
order unity distance from the last KAM curve. Then, the points
$(x_1,y_1)=\Phi(\xi_1= \xi_2, \eta_1=c_1/\xi_2)$ and $(x_2,y_2)$
have order unity distance, despite the fact that $(\xi_1,\eta_1)$
and $(\xi_2,\eta_2)$ are $\delta$-close.

For $\delta\rightarrow 0$, the image under $\Phi$ of a disc $\Delta$
of radius $\delta$ centered around some point $(\xi_0,\eta_0)$ in
the plane $(\xi,\eta)$ is an ellipse in the plane $(x,y)$ given by
the quadratic form
\begin{equation}\label{ellipse}
(\Delta x,\Delta y) M^TM(\Delta x,\Delta y)^T = \delta^2
\end{equation}
where $\Delta x = x-x_0$, $\Delta y=y-y_0$, with $(x_0,y_0) =
\Phi(\xi_0,\eta_0)$, and $M$ is the Jacobian matrix of the
transformation $\Phi$ evaluated at $(\xi_0,\eta_0)$. In fact, for
$\delta\rightarrow 0$ the linear approximation is valid. The
semi-axes of the ellipse have size $\delta/\sqrt{\mu_1}$,
$\delta/\sqrt{\mu_2}$, where $\mu_1,\mu_2$ are the eigenvalues of
$M^TM$. The major semi-axis is $s=\delta/
\sqrt{{min(\mu_1,\mu_2)}}$. For a fixed hyperbola $c=c_0$, the
quantity $S=1/ \sqrt{min(\mu_1,\mu_2)}$ can be expressed as a
function of $\xi_0$ only. If we set $c_0=c_{crit}$, we can express
the quantity $s$ as a function of the co-ordinate $\xi_0$ along the
curve $c=c_0$, with $\delta=\epsilon/\xi_0$ chosen so that the disc
is limited between the hyperbolae $c=c_1$ and $c=c_2$. We readily
find
\begin{equation}\label{sxi}
s(\xi_0,\epsilon) = {\epsilon S(\xi_0)\over \xi_0} .
\end{equation}
Equation (\ref{sxi}) implies that, for $\epsilon$ arbitrarily small,
the distance $s$ can become arbitrarily large, provided that
$S(\xi_0)$ grows with $\xi_0$ {\it faster than linearly}. Figure
\ref{fig19} shows a numerical computation indicating that this is
indeed the case in our numerical examples. Figure \ref{fig19}a shows
the image, under $\Phi$, of a disc of size $\delta \approx 2.6\times
10^{-6}$, centered around the point $\xi_0=7600$, $\eta_0=c_0/
\xi_0$, with $c_0=c_{crit}=0.41$. Then, $\epsilon=0.02$, i.e., the
uppermost and lowermost points of the disc touch the hyperbolae
$c=0.43$ and $c=0.39$ respectively. We see that the image
$\Phi(\Delta)$ is a very elongated ellipse whose most distant point
from the center has a  distance of order unity from the last KAM
curve.

In order to compute the matrix $M$ for the point $(\xi_0,\eta_0)$,
we use a similar procedure as in section 2. Namely, we first compute
the pre-image of $\xi_0$ on the $(\xi,\eta)$ plane by making a back
transform for a number of $n$ iterations using Eq.(\ref{unf}) so as
to find the mapping $\xi$ of $\xi_0$ in the central region of
Fig.\ref{fig02}b. This value of $\xi$ is of order unity. Then we can
apply the truncated series $\Phi$  in order to find the
corresponding $(x_0,y_0)$ point in the $(x,y)$ plane. Finally we
compute the $N-th$ forward iterate of the numerical map
(\ref{henmap}).  The matrix $M$ is computed as the product of all
the Jacobian matrices of the intermediate steps. This allows the
computation of $S(\xi_0)$.

Repeating the process for $n$ consecutive pre-images of
$(\xi_0,\eta_0)$ allows to obtain a numerical estimate of the form
of the function $S(\xi_0)$. As shown in Fig.\ref{fig19}b, this is
fitted by a power-law of the form $S(\xi_0)\sim S_0\xi_0^{p}$, with
constant $S_0=3.9$ and exponent $p\approx$1.4. Thus, $S(\xi_0)$
grows with $\xi_0$ faster than linearly.

We now show that the above analysis allows to estimate a {\it
minimum stickiness time} for the points $A',~C'$ of the initial
Moser arc of the hyperbola $c_2=c_0-\epsilon$. Setting
$\xi_0=c_0^{1/2} \Lambda(c)^N$ (for the N-th image of the point
$A'$), one has that $S(\xi_0)$ becomes equal to $S(\xi_0)=1$ after a
minimum number of iterations
\begin{equation}\label{nmin}
N_{min}=-{\log\epsilon+0.5(p-1)(\log c_0)-\log S_0\over
(p-1)\log\Lambda(c_0)}~~.
\end{equation}
Note, that $p>1$ is a necessary condition for the stickiness effect,
since only then $N\rightarrow\infty$ as $\epsilon\rightarrow 0$.

Figure \ref{fig19}c shows a comparison between numerically obtained
stickiness times (black curve) for initial conditions
$\xi_0=\eta_0=c^{1/2}$ with $c=c_{crit}-\epsilon$, as a function of
$\epsilon$, along with the prediction of formula (\ref{nmin}) (red
curve) for the minimum stickiness time. We observe that the
prediction actually represents the true stickiness time for
$\epsilon>0.02$. On the other hand, for $\epsilon<0.02$ the true
stickiness time is substantially larger than the minimum stickiness
time. In fact, this distinction is related to the existence of an
`inner' and 'outer' stickiness zone (see Contopoulos et al. 1997,
Efthymiopoulos et al. 1997). In the inner stickiness zone, the
stickiness time increases exponentially with $1/\epsilon$, thus it
is much larger than the estimate of Eq.(\ref{nmin}). We thus
conclude that the estimates of the theory of invariant Moser curves
predict qualitatively the stickiness phenomenon, but provide a
useful quantitative estimate of the stickiness time only in the
outer stickiness zone around an island of stability.

\section {Periodic orbits}
 For particular values of $c$, some self-intersections of the
invariant Moser curves correspond to periodic orbits. As noted
already by Birkhoff (1927, 1935), as $c$ goes to zero, the periodic
orbits lying on curves of smaller and smaller $c$ accumulate to
one or more homoclinic points, i.e., the points of intersection
of the curves $c=0$.
\begin{figure}
\centering
\includegraphics[scale=0.4]{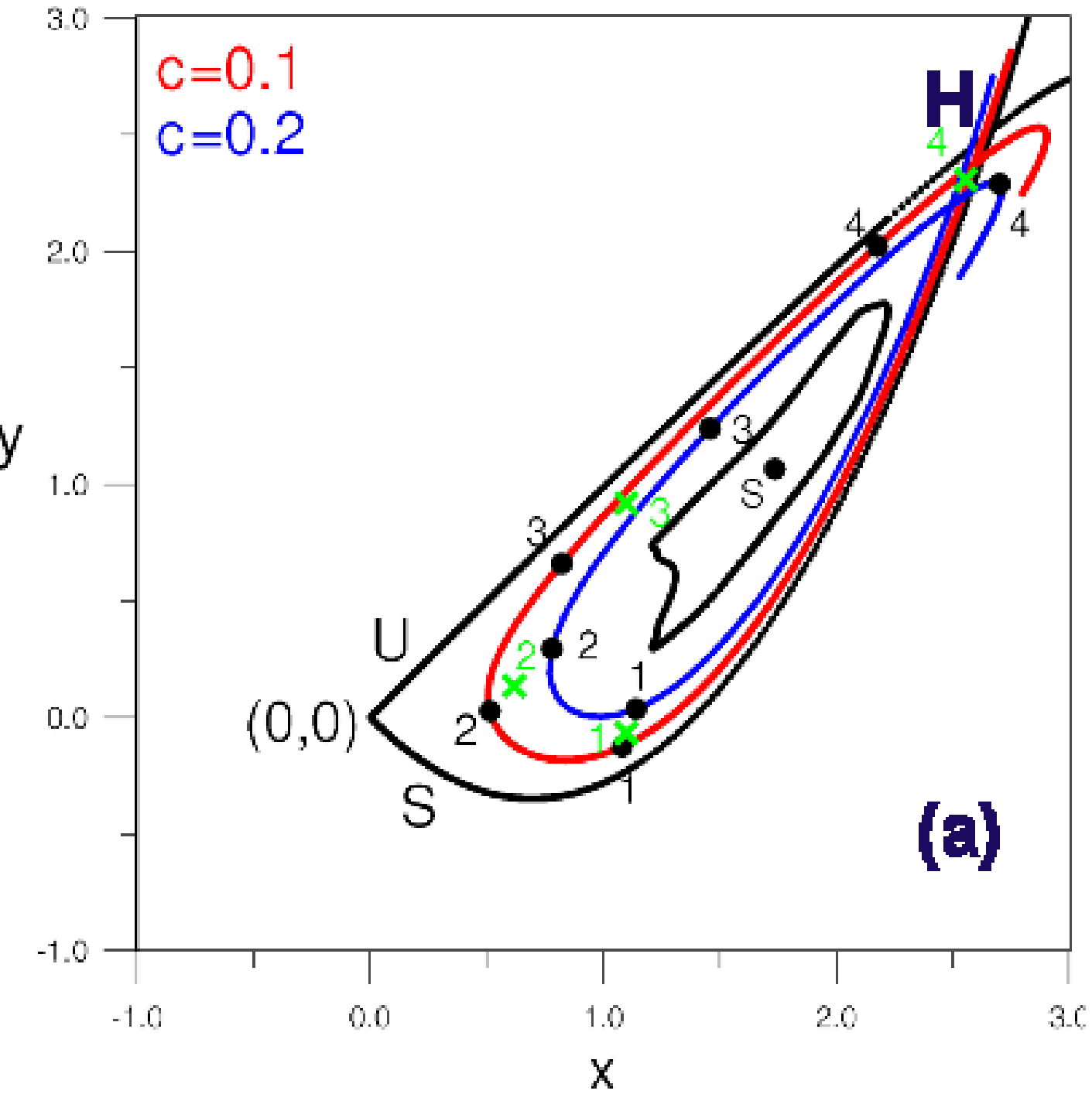}
\includegraphics[scale=0.4]{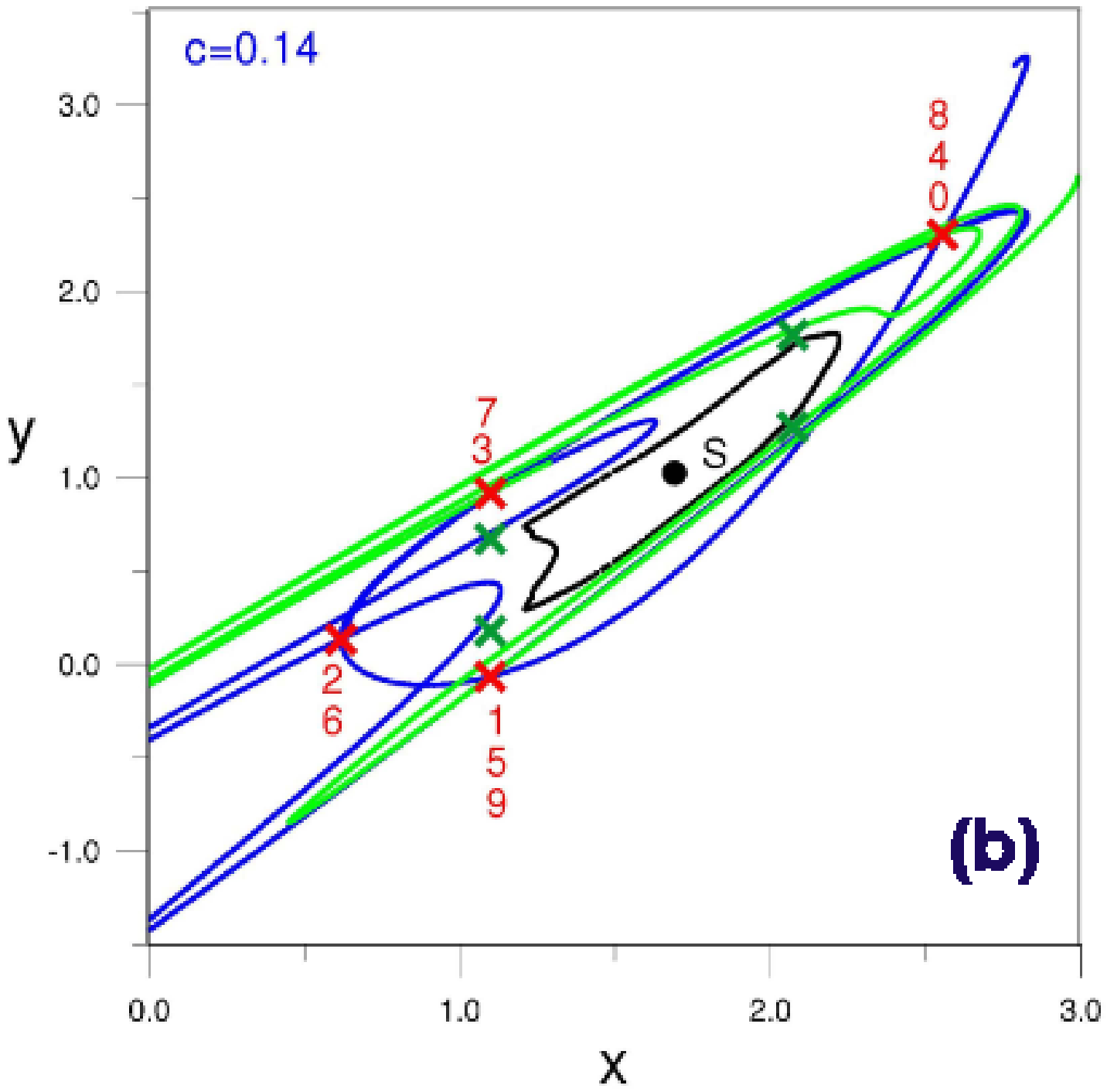}
\includegraphics[scale=0.37]{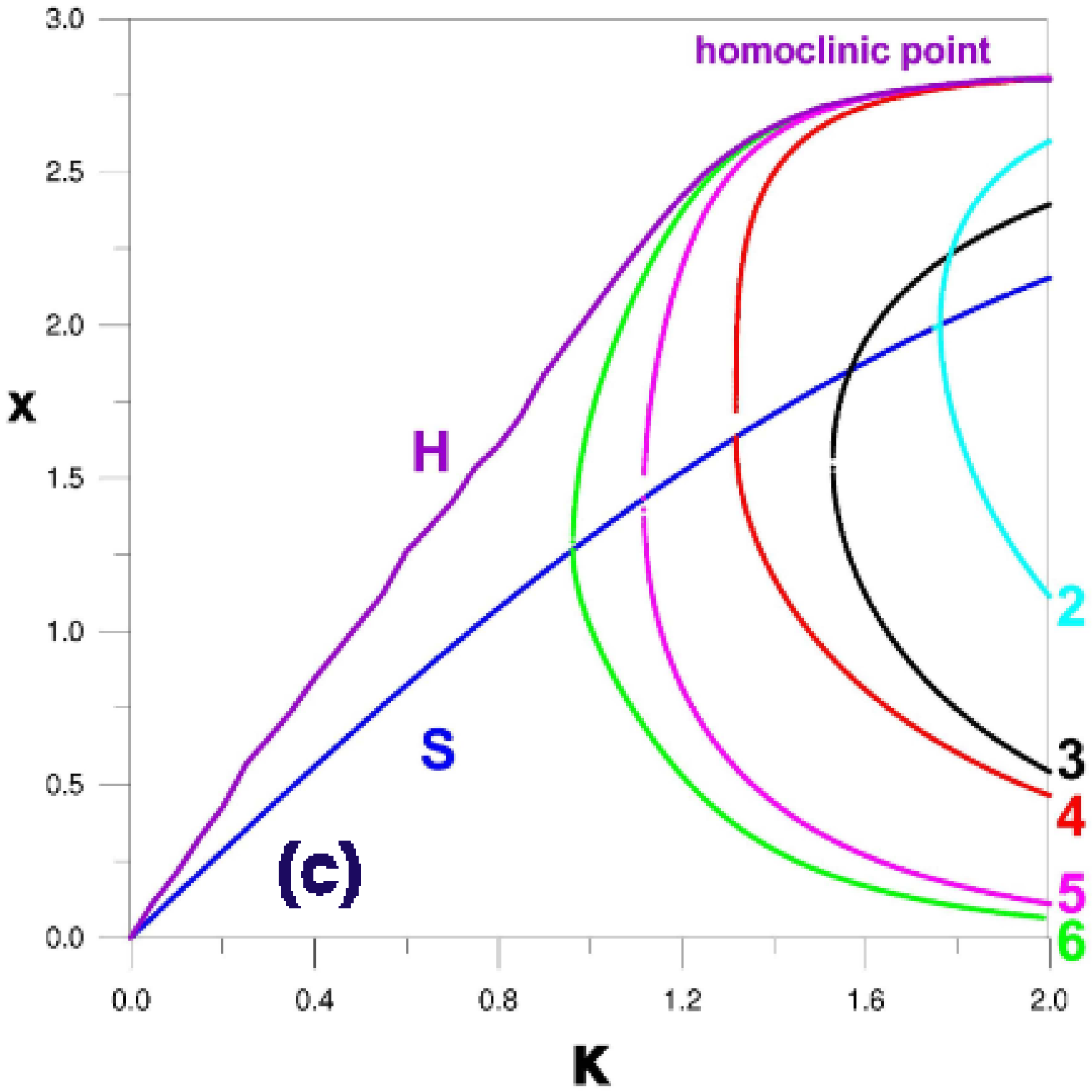}
\includegraphics[scale=0.45]{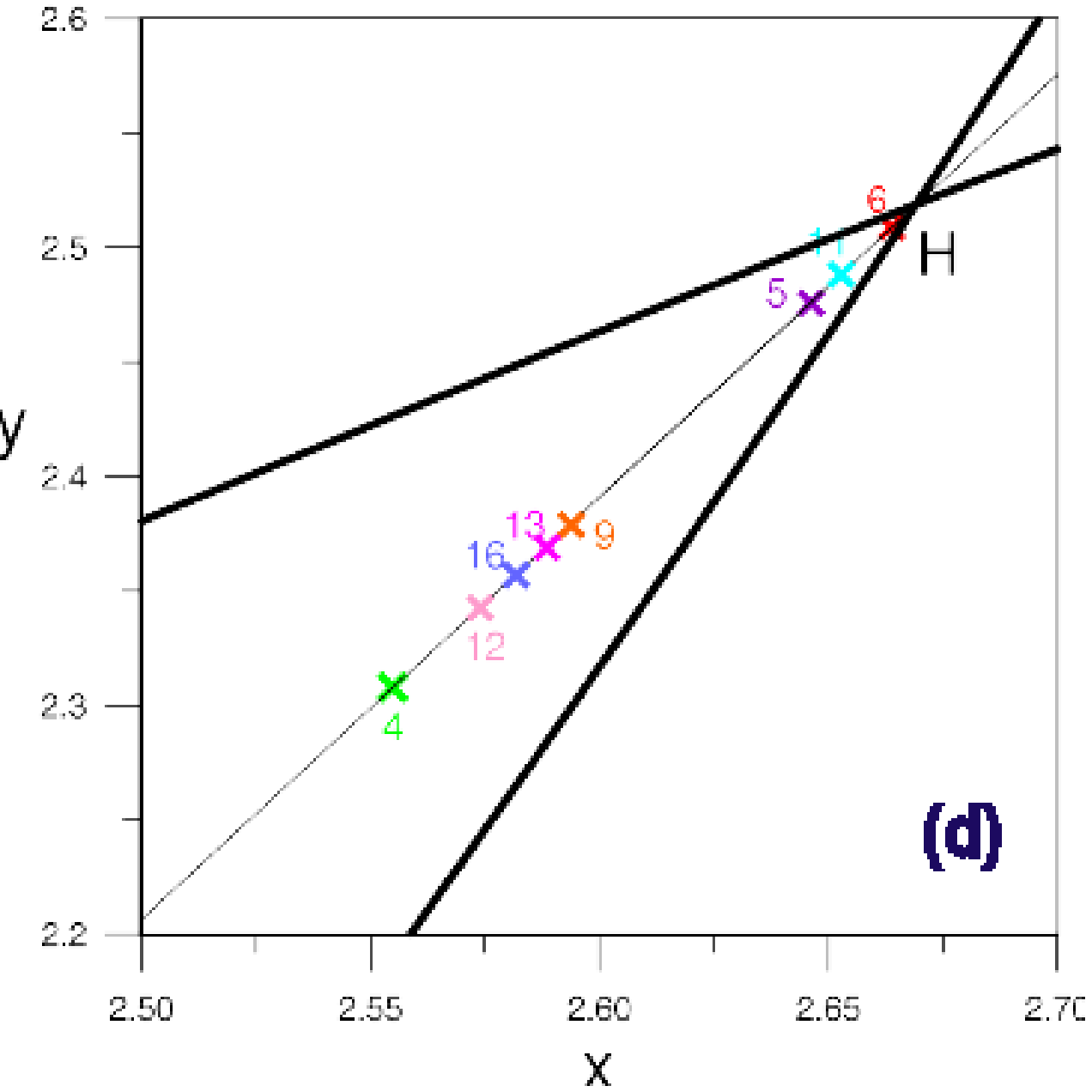}
\caption{ (a) The asymptotic curves from the origin ($U$ unstable
and $S$ stable) (black), the invariant curves $c=0.1$ (red) and
$c=0.2$ (blue) and the last KAM curve around the stable periodic
point $S$. The successive iterations of the intersection points on
the curves $c=0.1$ and $c=0.2$ are marked with numbers from 1 up to
4. Between these two curves an unstable periodic orbit of
multiplicity 4 lies on a curve $c=0.14,$ and is marked by green
crosses (b) The invariant curve $(c=0.14)$ (blue) passing through
the $4$ points $0,1,2,3$ (red crosses) of the periodic orbit of
period $-4$(clockwise), This curve continues (as blue) through the
point 4(=0), 5(=1), 6(=3), 7(=3), and then (as green) through 8(=0),
9(=1) etc. The second periodic orbit (green crosses) does not belong
to this curve. (c) Characteristics of the central periodic orbit
(blue curve) and of the families with rotation numbers $1/2$, $1/3$,
$1/4$, $1/5$ and $1/6$ bifurcating from it. The coordinate $x$ of
each periodic orbit is given as a function of the parameter
$\kappa$. The position of the first homoclinic point $H$
(intersection of the stable and unstable asymptotic curves) is
marked as well. The various characteristics approach asymptotically
the homoclinic curve as $\kappa$ increases (d)Periodic orbits close
to the homoclinic point $H$ for $\kappa=1.43.$ All these orbits are
close to a straight line and they are all regular.} \label{fig12}
\end{figure}
Da Silva Ritter et al. (1987) computed numerically a number of
unstable periodic orbits close to homoclinic points. An example of
their method is shown in Fig.\ref{fig12}a. We construct the
invariant curves with $c=0.2$ (blue curve) and $c=0.1$ (red curve),
up to points beyond their first self intersection, together with the
case $c=0$ (black curve) that corresponds to the stable and unstable
asymptotic curves of the unstable periodic orbit (0,0). The points
of intersection are
 numbered $0$ on each curve. Then we find the successive
iterates, using the map (\ref{henmap}), of the intersection points
$0$ along each curve. We see that along the curve $c=0.2$ the points
$0,1,2,3,4$ form more than one rotation around the center of the
island S, while, along the curve $c=0.1$ the fourth iterate of $0$
does not complete one full rotation. Therefore between the curves
$c=0.2$ and $c=0.1$ there is an invariant curve along which the point
$4$ coincides with the original point~$0$. Expressing this last
condition algebraically allows to compute the value of $c$ by
root-finding. We find $c\simeq 0.14$. Then, the point of intersection
(0 or 4) is also found numerically. The four points $0,1,2,3$ represent
a periodic orbit of period-4, shown by green crosses. The orbit has
rotation number $1/4$ around S.

If we continue the invariant curve for $c=0.14$ beyond the first
passage from the point 0 (or 4) of self-intersection, the curve
intersects itself at many more points. Most of these
self-intersections do not define periodic points (Fig.
\ref{fig12}b). However, the curve passes also infinitely many times
from the points 0,1,2,3 of the period-4 orbit (shown here with red
crosses). In Fig.\ref{fig12}b we see also a second period-$4$
periodic orbit (dark green crosses) which is unstable. This periodic
orbit does not lie on the same invariant curve with $c=0.14$, but it
belongs to a different Moser curve.

In a similar way we find periodic orbits of rotation number 1/5,
1/6, etc. It is obvious that there are infinitely many periodic
orbits as we approach closer to the homoclinic point H
(Fig.\ref{fig12}c) of periods $n=4,5,6...$ having smaller and
smaller rotation numbers around the stable orbit $S$. The homoclinic
orbit itself can be considered as a limiting periodic orbit with
infinite period and zero rotation number. All these orbits are
generated by pitchfork bifurcations from the central periodic orbit,
by varying the parameter $\kappa$ of the mapping (\ref{henmap}).
Figure \ref{fig12}c shows the characteristic curves of the periodic
orbits with $n=2$ to $6$, along with the characteristic curve of the
central stable orbit $S$. The position of the homoclinic point $H$
as a function of $\kappa$ is shown in the same plot, showing clearly
how orbits of lower and lower $n$ approach closer and closer to the
homoclinic point H as $\kappa$ increases.

Besides the periodic orbits with $rot=1/n$, there are periodic
orbits with $rot=2/n, 3/n$ etc. Such orbits were found by da Silva
Ritter et al. (1987) by using the intersections of two or more Moser
invariant curves. Between the orbits $1/n$ and $1/(n+1)$ we find
orbits forming Farey trees, i.e. the orbits $2/(2n+1)$, $3/(3n+1)$,
$3/(3n+2)$, etc. For $\kappa$ sufficiently large, all these orbits
are unstable and accumulate to the homoclinic point H. As shown in
Fig. \ref{fig12}d, the points of intersection of the corresponding
Moser curves are delineated approximately along a straight line.

A main empirical remark is now the following: we have checked that
{\it all} the periodic orbits found by the intersections of the
invariant Moser curves of the central unstable point $O$ in the
model (\ref{henmap}) belong to the above category, i.e., they are
generated by bifurcations from the stable central periodic orbit S
and they form Farey trees as indicated above. In fact, all the
periodic orbits given in Tables II and III of da Silva Ritter et al.
(1987) belong also to the same category. Such orbits, generated by
bifurcations from an initial orbit are called {\it regular} (see
Contopoulos 2002). It is of interest to note that in this particular
problem the lobes of the invariant manifolds of the central unstable
point $O$ cannot intersect themselves, thus they cannot generate
irregular periodic orbits. This is due to the fact that on the left
of the central unstable point $O$ the asymptotic curves extend to
infinity and the lobes on the right of these asymptotic curves
cannot come back to intersect themselves.

On the other hand, in other cases where the asymptotic curves do not
extend to infinity, there are irregular periodic orbits inside
intersecting lobes (Contopoulos and Polymilis 1996, Contopoulos and
Grousouzakou 1997). Thus, the possibility remains open that in other
cases some intersections of one or more invariant Moser curves
represent irregular periodic orbits. This subject is left for future
study.

\section {Conclusions}
The Birkhoff-Moser normal form for a 2D symplectic mapping yields an
integrable approximation of the mapping around an unstable
equilibrium point, represented by a convergent series in a domain
around the origin. Inside the domain of convergence, the dynamics
takes, in new variables $(\xi,\eta)$, a simple form of a hyperbolic
map: $\xi'=\Lambda(c)\xi$, $\eta'=\Lambda(c)^{-1}\eta$, where
$c=\xi\eta=\xi'\eta'$ labels invariant hyperbolae. The function
$\Lambda(c)$ is given as a series in powers of $c$, while the
mapping in the original variables $(x,y)$ can be represented by a
normalizing canonical transformation $(x,y)=\Phi(\xi,\eta)$. We
studied in detail the convergence properties of the series $\Phi$
and $\Lambda$ in an example of the 2D symplectic H\'{e}non map. The
invariant manifolds of the unstable point $(x,y)=(0,0)$ are given by
the images, under $\Phi$, of the axes $\xi=0$, $\eta=0$, i.e. $c=0$.
However, in the present paper we focus on the properties of the
curves corresponding to $c\neq 0$. In particular, we study the forms
of these curves, as well as what they imply for the structure of
chaos in the domain around the origin. Our main results are the
following:

1) By using a simple numerical criterion of absolute convergence, we
find numerical evidence that the limit of the domain of convergence
of both series $\Lambda$ and $\Phi$ is close to a hyperbola with
$c=c_{max}$, where, in our numerical example, we find $c_{max}\simeq
0.49$. We provide some estimates of the behavior of truncation
errors, and propose an `extrapolation' technique allowing to better
determine the limit of the domain of convergence even when the
truncation error of the series is relatively large.

2) The truncation error of the series is not uniform along an
invariant curve with $c<c_{max}$. In general, at a fixed truncation
order $r$, the error is minimum for the part of the hyperbola near
the diagonal $\xi=\eta$, while it increases exponentially as
$\xi\rightarrow\pm\infty$ or $\eta\rightarrow\pm\infty$. Thus, in
order to accurately represent a segment of the hyperbolae of length
$s$ around the diagonal in the original variables using the series
$\Phi$, one needs to reach a truncation order $r\sim\exp(s)$. This
implies the consistency of the series with the fact that they
represent chaotic motions. Namely, in order to accurately determine
more and more iterates of a chaotic orbit using the series, one
needs to compute an exponentially increasing number of series terms.

3) Even so, one is able to obtain the form of the curves of constant
$c$ in the original variables ($x,y$) using a conjugation of the
series with the mapping equations, as proposed in da Silva Ritter et
al. (1987). In this way, we find the form of the domain of
convergence as mapped in the original variables, as well as many
other interesting properties of the Moser invariant curves. In
particular, we find that beyond a critical value $c=c_{crit}\simeq
0.41$, the curves lie entirely within an island of stability around
a stable periodic point $S$, at a distance from the (unstable)
origin. The fact that we can have invariant curves of this type
inside an island of stability does not contradict the existence of
KAM curves. In fact, the curves of constant $c$ represent the
phase-mixing taking place along {\it open sets} of initial
conditions both on KAM curves (regular) or on thin chaotic layers
between the KAM curves (chaotic). However, the truncation error
along the Moser invariant curves is non-uniform while along the KAM
curves it is uniform.

4) There are also curves $c=const$ which lie entirely outside the
last KAM curve of the island of stability, as well as curves
$c=const$ which intersect the last KAM curve. We compute the last
KAM curve by a numerical criterion of `rotation and twist angles'
(Voglis and Efthymiopoulos 1998). We also consider the problem of
how neighborhoods in the plane $(\xi,\eta)$ are mapped into
neighborhoods in the plane $(x,y)$ for hyperbolae arbitrarily close
(above and below) to the critical value $c=c_{crit}$. We demonstrate
that the properties of the normalizing transformation $\Phi$ {\it
imply} a stickiness effect. Namely, initial conditions close to the
diagonal for a curve with $c=c_0-\epsilon$, with $\epsilon>0$
arbitrarily small, have a minimum stickiness time increasing with
$\epsilon$ as $\sim-\log(\epsilon)$.

5) For particular values of $c$, the curves $c=const$ have
self-intersections corresponding to periodic orbits of a predictable
multiplicity. We compute such orbits, and, going to higher and
higher multiplicity, we find how they accumulate close to homoclinic
points of the invariant manifolds of the unstable fixed point at the
origin. We find the bifurcation histories of these periodic orbits
and show that they are all {\it regular}, i.e. belonging to Farey
trees built by bifurcations around the stable fixed point at the
center of the island of stability. In fact, taking into account the
fact that in our mapping model some initial conditions lead to
escapes, one can show that no irregular periodic orbits can be
generated by the lobes of the invariant manifolds of the central
unstable fixed point.

6) Finally, we emphasize the fact that, all the points of a chaotic
orbit near the unstable periodic orbit of the origin (0,0), although
they seem to be distributed randomly, they belong in fact to
particular "Moser invariant curves". Therefore, these curves define
a "structure of chaos", which can be computed by analytical formulae
inside the region of convergence.

\section*{Acknowledgments}
 This work has been completed in the frame of the research project
 of RCAAM (Research Center for Astronomy and Applied Mathematics) "Analytic computation of invariant manifolds and the
structure of chaos".

\end{document}